\documentclass[journal]{IEEEtran}
\usepackage{cite}
\usepackage[pdftex]{graphicx}
\usepackage{epsf}
\usepackage{epstopdf}
\usepackage{graphicx}
\usepackage[cmex10]{amsmath}
\usepackage{eqnarray}
\usepackage[tight,footnotesize]{subfigure}
\usepackage{enumerate}
\usepackage{algorithm}
\usepackage{algpseudocode}
\usepackage{authblk}
\usepackage{tikz}
\usepackage{epic}

\begin{document}

		\title{Applications of Compressed Sensing in Communications Networks}
		\author[1]{Hong Huang}
        \author[2]{Satyajayant Misra}
        \author[1]{Wei Tang}
        \author[1]{Hajar Barani}
        \author[1]{Hussein Al-Azzawi}
        \affil[1]{Klipsch School of Electrical and Computer Engineering, New Mexico State University, NM, USA}
        \affil[2]{Department of Computer Science, New Mexico State University, NM, USA}

	\maketitle
	\begin{abstract}
This paper presents a tutorial for CS applications in communications networks. The Shannon's sampling theorem states that to recover a signal, the sampling rate must be as least the Nyquist rate. Compressed sensing (CS) is based on the surprising fact that to recover a signal that is sparse in certain representations, one can sample at the rate far below the Nyquist rate. Since its inception in 2006, CS attracted much interest in the research community and found wide-ranging applications from astronomy, biology, communications, image and video processing, medicine, to radar. CS also found successful applications in communications networks. CS was applied in the detection and estimation of wireless signals, source coding, multi-access channels, data collection in sensor networks, and network monitoring, etc. In many cases, CS was shown to bring performance gains on the order of 10X. We believe this is just the beginning of CS applications in communications networks, and the future will see even more fruitful applications of CS in our field.
	\end{abstract}


	\section{Introduction}
		\label{section:intro}
		\IEEEPARstart{P}{rocesssing} data is a big part of modern life. Interesting data typically is sparse in certain representations. An example is an image, which is sparse in, say, the wavelet representation. The conventional way to handle such signal is to acquire all the data first and then compress it, as is done in image processing. The problem with this kind of processing is, as Donoho puts it in his seminal paper \cite{donoho2006}: "Why go to so much effort to acquire all the data when most of what we get will be thrown away? Can we not just directly measure the part that will not end up being thrown away?" Indeed, compressed sensing (CS) does just that-- measuring only the part of data that is not thrown away. The well-known Shannon's sampling theorem states that to recover a signal exactly, the sampling rate must be as least the Nyquist rate, which is twice the maximum frequency of the signal. In contrast, using CS, far fewer samples or measurements at far below the Nyquist rate are required to recover the signal as long as the signal is sparse and the measurement is incoherent, the exact meanings of which will be revealed later. In addition, CS has other attractive attributes, such as universality, fault-tolerance, robustness to noise, graceful degradation, etc.

Since its introduction in \cite{candes2006} and \cite{donoho2006} in 2006, CS has received much attention in the research community. Thousands of papers have been published on topics related to CS, and hundreds of conferences, workshops and special sessions have been devoted to CS \cite{duarte2011}. CS has been showed to bring significant performance gains in wide-ranging applications from astronomy, biology, communications, image and video processing, medicine, to radar. Although there are good survey and tutorial papers on CS itself \cite{candes2006a, Baraniuk2007, candes2008, davenport2012}, there is no good tutorial paper on the application of CS in communications networks to the best of the authors' knowledge. We hope this paper will close the gap. Our intention is to enable the readers to implement their own CS applications after reading this tutorial. Therefore, instead of covering a wide-range of CS applications, we focus on a few representative applications and provide very detailed description for the covered applications. The intended readers should have some basic backgrounds in signal processing, communications networks in general, and sensor networks in particular. This paper contains a fair amount of mathematical formulas, since CS is a mathematical tool and mathematics is best expressed by formulas. Although CS has been applied in some areas of communications networks, we believe deeper and wider applications of CS in communications networks are plausible. Readers can use this tutorial to familiarize themselves with CS and seek a wider-range of fruitful applications.

CS can be applied in various layers of communications networks. At the physical layer, CS can be used in detecting and estimating sparse physical signals such as ultra-wide-band (UWB) signals, wide-band cognitive radio signals, and MIMO signals. Also, CS can be used as erasure code. At the MAC layer, CS can be used to implement multi-access channels. At the network layer, CS can be used for data collection in wireless sensor networks, where the sensory signals are usually sparse in certain representations. At the application level, CS can be used to monitor the network itself, where network performance metrics are sparse in some transform domains. In many cases, CS was shown to bring performance gains on the order of 10X. We believe this is just the beginning of CS applications in communications networks, and the future will see even more fruitful applications of CS in our field.

The rest of the paper is organized as follows. In Section II, we provide an overview of the CS theory. In Section III, we describe the algorithms that implement CS. In Section IV, we describe some variations of the CS theory. In section V, we provide a tutorial on CS applications in the physical layer. In Section VI, we described CS applications as erasure code and in the MAC layer. In section VII, we describe how CS is used in the network layer. In Section VIII, we describe CS applications in the application layer. We conclude in Section IX.

	\section{The Theory of Compressed Sensing}
		\label{section:compressed_sensing}
		The theory of CS was mostly established in \cite{candes2006} and \cite{donoho2006}, where the reader can find proofs of the results described below.
\subsection{Sparse Signal}
	For the vector $x = (x_i), i=1, 2,...,n$ its $l_p$ norm is defined by
		\begin{equation}
		\label{eqn:p_norm}
			\| x \|_p = (\sum_{i = 1}^{n} |x|^p)^{1/p}.
		\end{equation}
The above definition applies to either row or column vector. Note that for $p = 0, \| x\|_0$ is the number of nonzero elements in $x$; for $p = 1, \| x\|_1$ is the summation of the absolute values of elements in $x$; for $p = 2, \| x\|_2$  is usual Euclidean norm; and for $p = \infty, \| x\|_\infty$  is the maximum of the absolute values in $x$. We call a signal $k$-sparse if $\| x\|_0 \leq k$, namely $x$ has only $k$ nonzero elements. We also say the sparsity of the signal is $k$.

In fact, few real-world signals are exactly $k$-sparse. Rather, they are compressible in the sense that they can be well-approximated by a $k$-sparse signal. Denote $\Sigma_k$ is the set of all $k$-sparse signals. The error incurred by approximating a compressible signal $x$ by a $k$-sparse signal is given by
        \begin{equation}
		\label{eqn:error_sigma}
			\sigma_k(x)_p = \underset{\hat{x} \in \Sigma_k}{\text{min}}   \|\hat{x} - x \|_p.
		\end{equation}
If $x$ is $k$-sparse, $\sigma_k(x)_p = 0$. Otherwise, the optimal $k$-sparse approximation of $x$ is the vector that keeps the $k$ largest elements of $x$ with the rest elements setting to zero.

Consider a compressible signal $x$ and order its elements in descending order so that $|x_1|\geq|x_2|\geq...\geq|x_n|$. We say the elements follow a power-law decay if
        \begin{equation}
		\label{eqn:decay_power}
			|x_i| \leq c i^{-\alpha}
		\end{equation}
where $c$ is a constant and $\alpha > 0$ is the decay exponent. If $x$ follows a power-law decay, there exist some constants $c, \alpha > 0$, such that \cite{devore1998}
        \begin{equation}
		\label{eqn:error_sigma2}
			\sigma_k(x)_2 \leq ck^{-\alpha - 1/2}.
		\end{equation}
In other words, the error also follows a power-law decay.

\subsection{The Basic Framework of CS}
	Although CS applies to both finite and infinite dimensional signals, we focus on finite dimensional signals here for the ease of exposition. Consider an $n$-dimensional signal $f$, which has a representation in some orthonormal basis $\Psi = [\psi_1, \psi_2, ...,\psi_n]$ where
		\begin{equation}
		\label{eqn:f_rep}
			f = \sum_{i = 1}^{n} x_i \psi_i = \Psi x.
		\end{equation}
In the above, $x_i$ and $\psi_i$ (a column vector) are the \emph{i}th coefficient and \emph{i}th basis, respectively.
The CS theory says that if $f$ is sparse in the basis $\Psi$, which needs not be known \emph{a priori}, then, under certain conditions, taking $m$ nonadaptive measurements of $f$ suffices to recover the signal exactly, , where $m \ll n$. Each measurement $y_j$ is a projection of the original signal, i.e., the $m$-dimensional measurement can be represented by
		\begin{equation}
		\label{eqn:y_meas}
			y = \Phi x
		\end{equation}
where $y$ is the measurement vector and $\Phi$ is an $m\times n$ sensing matrix. Since $m \ll n$, to recover $x$ from $y$ is an ill-posed inverse problem. In other words, there might be multiple $x'$s that satisfy (\ref{eqn:y_meas}). However, we can take advantage of the fact that $x$ is sparse in a certain representation $\Psi$, and formulate the following optimization problem:
		\begin{equation}
		\label{eqn:p_0}
			\text{($P_0$)}  \quad \underset{x}{\text{min}}  \|\Psi x\|_0 \quad \text{subject to}\quad y = \Phi x.
		\end{equation}

It is known that solving $P_0$ is NP-hard. Fortunately, it was shown that one can replace the $l_0$ norm by $l_1$ norm, and formulate the following optimization problem instead as \cite{candes2006b, donoho_tanner2005, candes2005}
		\begin{equation}
		\label{eqn:p_1}
			\text{($P_1$)}  \quad  \underset{x}{\text{min}}  \|\Psi x\|_1 \quad \text{subject to}\quad y = \Phi x.
		\end{equation}
It can be shown if the signal is sufficiently sparse, the solutions to $P_0$ and $P_1$ are the same \cite{candes2006b}. $P_1$ is a convex linear-programming problem with efficient solution techniques.

In the rest of the section, we will describe what kind of sensing matrix will enable signal recovery, and how many measurements are needed for the recovery. The most well-known answers to the above questions are expressed in terms of spark, mutual coherence, and restricted isometry property.

\subsection{Spark}
The spark of a matrix $\Phi$ is the smallest number of columns of $\Phi$ that are linearly dependent \cite{donoho_elad2003}. To see why spark is relevant, consider the $k$-sparse solution to the linear equation $y = \Phi x$. The solution is not unique if there are two different $k$-sparse vectors $x$ and $x'$ such that $y = \Phi x = \Phi x'$, which implies $\Phi (x - x') = 0$. Equivalently, $(x - x')$ is in the null space of $\Phi$. Since the sparsity of $(x - x')$ is at most $2k$, so to ensure the uniqueness of the solution, the null space of $\Phi$ can not have vectors whose sparsity is less or equal to $2k$. In other words, to guarantee the uniqueness of the $k$-sparse solution, the smallest number of linear dependent columns of $\Phi$ must be larger than $2k$, or
        \begin{equation}
		\label{eqn:spark}
			\text{spark}(\Phi) > 2k.
		\end{equation}

For the $m\times n$ matrix $\Phi$, $\text{spark}(\Phi) \in [2, m + 1]$. So, to make the solution unique, the number of measurements taken must satisfy
        \begin{equation}
		\label{eqn:spark_meas}
			m \geq 2k.
		\end{equation}

It turns out that equation (\ref{eqn:spark}) is also the sufficient condition to guarantee the error bound given by \cite{cohen2009}
        \begin{equation}
		\label{eqn:spark_error}
			\|\hat{x} - x \|_2 \leq c \sigma_k(x)_1 / \sqrt{k}
		\end{equation}
where $x$ and $\hat{x}$ are the original and the recovered signals, $c$ is a constant, and $\sigma_k(x)_1$ is defined in (\ref{eqn:error_sigma}). So, if $x$ is exactly $k$-sparse, then $\sigma_k(x)_1 = 0$, and the signal recovery is exact.
\subsection{Mutual Coherence Property (MIP)}
According to CS theory \cite{donoho_huo2001}, to enable signal recovery, the sensing matrix $\Phi$ must be incoherent with the sparse representation $\Psi$. For the ease of exposition, we assume the column vectors of both matrices form ortho-bases of $R^n$, which is not required by CS theory. Formally, the mutual coherence between two matrices is defined by \cite{donoho_elad2003, tropp2007}
		\begin{equation}
		\label{eqn:coherence}
			\mu(\Phi,\Psi) = \sqrt{n}\underset{1\leq k,j\leq n}{\text{max}}\mid \langle\phi_k, \psi_j\rangle\mid
		\end{equation}
where $\langle\phi_k, \psi_j\rangle$ is the inner product of $k$th column in $\Phi$ and $j$th column in $\Psi$. In other words, the coherence measures the maximum correlation between the columns of $\Phi$ and $\Psi$, and it has a range of $[1, \sqrt{n}]$ \cite{rosenfeld1996, strohmer2003, welch1994}. The coherence is large when the two matrices are closely correlated, and is small otherwise.

With probability $1-\delta$, a $k$-sparse signal can be recovered exactly if $m$ (the number of measurements) satisfies the following condition for some positive constant $c$ \cite{Baraniuk2007, candes2006c}
		\begin{equation}
		\label{eqn:bound-meas}
			m \geq c\mu^2(\Phi,\Psi)k\log (n).
		\end{equation}
Note that the number of measurements required increases linearly with the sparsity of the signal and quadratically with the coherence of the matrices. Note that the above MIP bound is not tight for some measurement matrices.
\subsection{Restricted Isometry Property (RIP)}
	In CS, an important concept is the so-called \emph{restricted isometry property}\cite{candes_tao2005, tropp2004}. We define the isometry constant $\delta_k$ of a matrix $\Phi$ as the smallest number such that the following holds for all k-sparse vectors $x$
		\begin{equation}
		\label{eqn:RIP}
			(1 - \delta_k)\|x\|^2_{2} \leq \|\Phi x\|^2_{2} \leq (1 + \delta_k)\|x\|^2_{2}
		\end{equation}

Informally, we say a matrix $\Phi$ holds RIP of order $k$ if $\delta_k$ is not too close to one. Three remarks follow:
\begin{itemize}
\item A transformation of $k$-sparse signal by a matrix holding RIP approximately preserves the signal's $l_2$ norm. The degree of preservation is indicated by $\delta_k$, with $\delta_k = 0$ being exactly preserving and $\delta_k = 1$ being not preserving. Because RIP preserves $l_2$ norm, for a matrix $\Phi$ holding RIP, its null space does not contain $k$-sparse vector. As mentioned earlier, this is important because for any solution $x$ of (\ref{eqn:p_1}), $x + x_0$ is also a solution, where $x_0$ is a vector in the null space of $\Phi$, making the solution not unique. RIP precludes such situation from arising.
\item RIP implies that all subsets of $k$ columns of $\Phi$ are nearly orthogonal.
\item A further implication of RIP is that if $\delta_{2k}$ is sufficiently small, then the measurement matrix $\Phi$ preserves the distance between any pair of $k$-sparse signals, i.e., $(1 - \delta_{2k})\|x_1 - x_2\|^2_{2} \leq \|\Phi (x_1 - x_2)\|^2_{2} \leq (1 + \delta_{2k})\|x_1 - x_2\|^2_{2}$.
\end{itemize}

When the measurement involves noise, we can revise $P_1$ by relaxing the measurement constraint as below
		\begin{equation}
		\label{eqn:p_2}
			\text{($P_2$)}  \quad  \underset{x}{\text{min}}  \|\Psi x\|_1 \quad \text{subject to} \quad \|y - \Phi x\|_2 \leq \epsilon
		\end{equation}
where $\epsilon$ bounds the amount of noise.
Using RIP, we can bound the error of signal recovery when noise is present. In fact, if $\delta_{2k} < \sqrt{2} - 1$, then for some constants $c_1$ and $c_2$ the solution $x^*$ to $P_2$ satisfies \cite{candes_tao2005, tropp2004}
		\begin{equation}
		\label{eqn:RIP-bound}
			\|x - x^*\|_2 \leq c_1\frac{\|x - x_k\|_1}{\sqrt{k}} + c_2\epsilon
		\end{equation}
where $x_k$ is the vector $x$ with all but the largest k elements set to zero. Two remarks follow:
\begin{itemize}
\item Consider the case where there is no noise. In such case, if $x$ is $k$-sparse, i.e., $x = x_k$, then the recovery is exact. On the other hand, if $x$ is not $k$-sparse, then the recovery is as good as if we know beforehand the largest $k$ elements and use them as an approximation.
\item The contribution of the noise is a linear term. In other words, CS handles noise gracefully.
\end{itemize}
\subsection{Relationships among Spark, MIP, and RIP}
Spark and MIP have the following relationship
        \begin{equation}
		\label{eqn:spark_mip}
			\text{spark}(\Phi) \geq 1 + 1/\mu(\Phi)
		\end{equation}
where $\mu(\Phi) = \mu(\Phi,\Phi)/ \sqrt{n}$. The above result is a straightforward application of the Gershgorin circle theorem \cite{gersgorin1931}.

RIP is strictly stronger than the spark condition. Specifically, if a matrix $\Phi$ satisfies the RIP of order $2k$, then $\text{spark}(\Phi) \geq 2k$.

RIP and MIP are related as follows. If $\Phi$ has unit-norm columns and coherence $\mu(\Phi)$, then $\Phi$ satisfies the RIP of order $k$ for all $k < 1/ \mu(\Phi)$.

For general matrix, it is hard to verify whether the matrix satisfies the spark and RIP conditions. Verifying MIP is much easier, since it involves calculating the inner products between columns of two matrices and taking the maximum of products, referring to (\ref{eqn:coherence}).
\subsection{Measurement Matrices}
Random matrices are commonly used for measurement matrices, though non-random matrices can also be used as long as they satisfy spark, MIP, or RIP requirements. By definition, random matrices are those that have elements following independent identical distributions. Examples of random matrices include those, whose elements follow Bernoulli distribution ($\text{Prob}(\phi_{i,j} = \pm 1/\sqrt{m}) = 1/2$) or Gaussian distribution with zero mean and variance of $1/m$.

With high probability, random matrices are incoherent with any fixed basis $\Psi$. Further, most random matrices obey the RIP and can serve as the measurement matrix $\Phi_{m\times n}$, as long as the following condition holds for some constant $c$ that varies with the particular matrix used \cite{davenport2010}
		\begin{equation}
		\label{eqn:condition_rip}
			m \geq ck\log{(n/k)}.
		\end{equation}

In fact, using random matrices is near-optimal in the sense that it is impossible to recover signal using substantially fewer measurements than the left-hand side of (\ref{eqn:condition_rip}).

	\section{The Algorithms for Implementing CS}
		\label{section:CS algorithms}
		There are three main types of algorithms for implementing CS: convex optimization algorithms, combinatorial algorithms, and greedy algorithms. Convex optimization algorithms require fewer measurements but are more computationally complex than those of combinatorial algorithms. Those two types of algorithms represent two extremes in the spectra of the number of measurements and the computational complexity. Greedy algorithms provide a good compromise between the two extremes, which we provide more detailed descriptions below. Table (\ref{table:CS_algorithms}) summarizes the computational complexity of various algorithms for implementing CS.
\begin{table*}[!ht]
\caption {Algorithms for implementing CS}
\begin{center}
\begin{tabular}{ l  p{5cm} }
\hline
Algorithms & Computation complexity of the best algorithm in the category   \\ \hline
Convex optimization algorithms \cite{candes2006, daubechies2004, figueiredo2007, grant, kim2007}                      &  $O(m^2 n^{3/2})$  \\
Greedy algorithms \cite{donoho2012, mallat1993, needell2009, needell2009a, needell2010, pati1993, rabbat2006}                &  $O(mn\log{k})$         \\
Combinatorial Algorithms  \cite{muthukrishnan2005, gilbert2007, cormode2009, iwen2010}    &  $O(k(\log{n})^c)$, where $c$ is a positive constant. \\ \hline
\end{tabular}
\end{center}
\label{table:CS_algorithms}
\end{table*}
\subsection{Convex Optimization Algorithms}
With convex optimization algorithms, we use an unconstrained version of $P_2$ given by the following
		\begin{equation}
		\label{eqn:p_4}
			\text{($P_4$)}  \quad  \underset{x}{\text{min}} \frac{1}{2} \|y - \Phi x\|^2_2 + \lambda \|\Psi x\|_1
		\end{equation}
where $\lambda$ can be selected based on how much weight we want to put on the fidelity to the measurements and the sparsity of the signal. Several ways to select $\lambda$ were discussed in \cite{eldar2009, friedman2010}. Efficient algorithms exist to solve the convex optimization problem, such as basis pursuit \cite{grant}, iterative thresholding that uses soft thresholding to set the coefficients of the signal\cite{daubechies2004}, interior-point (IP) methods that uses the primal-dual approach \cite{candes2006, kim2007}, and projected gradient methods that updates the coefficients of the signal in some preferred directions \cite{figueiredo2007, blumensath2008}.
\subsection{Greedy Algorithms}
Greedy algorithms iteratively approximate the original signal and its support (the index set of nonzero elements). Earlier algorithms include Matching Pursuit (MP) \cite{mallat1993}, and its improved version Orthogonal Matching Pursuit (OMP)\cite{pati1993, tropp2007}, which is listed below as a representative greedy algorithm.
		\begin{algorithm}
        	\caption{OMP Algorithm}
			\begin{algorithmic}[1]
\label{alg:omp}
				\Statex Input: $m\times n$ measurement matrix $\Phi = (\phi_i)$, where $\phi_i$ is the $i$th column of $\Phi$, $n$-dimensional initial signal $x$, and error threshold $\epsilon$.	
                \State Output: approximate solution $x_j$
				\State Set $j = 0$
				\State Set the initial solution $x_0 = 0$
				\State Set the initial residual $r_0 = y - \Phi x_0 = y$
				\State Set the initial support $S_0 = \emptyset$
				\Repeat

				\State Set $j = j + 1$
				\State Select index $i$ so that $\underset{i}{\text{max}} \|\phi^T_{i}r_j\|_2$, update $S_j = S_{j - 1}\cup{i}$
				\State Update $x_j = \underset{x}{\text{argmin}}\|\phi x - y\|_2$, subject to supp($x) = S_j$
				\State Update $r_j = (1 - P_j)y$, where $P_j$ denotes the projection onto the space spanned by the columns in $S_j$

				\Until{$\|r_j\|_2 \leq \epsilon$}
				\end{algorithmic}
		\end{algorithm}
		
In the algorithm listed above, we have set $\Psi$ to identity matrix for the ease of exposition. In the algorithm, we iteratively select the column of $\Phi$ that is most correlated with current residual $r_j$ and add it to the current support $S_j$ in step 7. Then, we update the current signal $x_j$ so that it is conforming to the measurements $y$ and current support $S_j$. Finally, we update the current residual $r_j$ so that it contains measurements excluding those included by the current support. Denote $\Phi_j$ as the sub-matrix containing only columns in the current support $S_j$, then $P_j = \Phi_j(\Phi^T_j \Phi_j)^{-1}\Phi^T_j$.
It can be shown that if the the signal $x$ is sufficiently sparse, i.e., $\|x\|_0 < \frac{1}{2}(1 + \mu(\Phi)^{-1})$, OMP can exactly recover $x$ when $\epsilon$ is set to zero\cite{tropp2004}. For noisy measurements, $\epsilon$ can be set to a nonzero value, for details of which the reader is referred to \cite{pati1993}. Compared with MP, OMP converges faster with no more than $n$ iterations.

Recent developments of this type of algorithms include the following. In \cite{needell2009a, needell2010}, regularized OMP (ROMP) is proposed that combines the speed and ease of implementation of the greedy algorithms with the strong performance guarantees of the convex optimization algorithms. In \cite{needell2009}, a method called compressive sampling matching pursuit (CoSaMP) is proposed, which provides the same performance guarantee as that of the best optimization-based methods, and requires only matrix-vector multiplies. In \cite{dai2009}, a subspace pursuit (SP) algorithm is proposed that has the same computational complexity as that of OMP but has the reconstruction accuracy on the same order as that of convex optimization algorithms. Both CoSaMP and SP incorporate ideas from convex optimization and combinatorial algorithms. In \cite{donoho2012}, Stagewise Orthogonal Matching Pursuit (StOMP) is proposed, which allows multiple coefficients of the signal to be added to the model in each iteration and performs a fixed number of iterations.

In Algorithm 2, we present CoSaMP \cite{needell2009} as an example of the state-of-art greedy algorithm. The notation we use is the following. For signal $x$, let $x_{(k)}$ denote the signal that keeps the $k$ largest components of $x$ while setting the rest of the components to zero. For an index set $T$, let $x_{|T}$ denote the signal that keeps the components in $T$ while setting the rest of the components to zero. Also, let $\Phi_T$ denote the column sub-matrix of $\Phi$ whose columns are listed in $T$. Finally, let $A^{\dagger} = (A^TA)^{-1}A^T$ denote the pseudo-inverse of the matrix $A$. The CoSaMP algorithm consists of five major steps. 1) Identification: We form a signal proxy $u$ from the current sample and identify the support $\Omega$ of the $2k$ largest components. 2) Support merger: We merge $\Omega$ with the support of the sparse signal in the previous round. 3) Estimation: We estimate the signal using the least-squares method to approximate the target signal on the merged support. 4) Pruning: We prune the resulting signal to be $k$-sparse. 5) Sample update: We update the sample $z$ so that it reflects the residual $r$ that contains the signal that has not been approximated. We stop when the halting criterion is met.

There are three major approaches to halting the algorithm. In the first approach, we halt the algorithm after a fixed number of iterations. In the second approach, the halting condition is $\|z\|_2 \leq \epsilon_1$. In the third approach, we can use $\|u\|_{\infty}$ to bound the residual $\|r\|_{\infty}$, and require that $\|r\|_{\infty} \leq \epsilon_2$. Using different approaches and parameters involve the tradeoff between approximation accuracy and computational complexity.

		\begin{algorithm}
        	\caption{CoSaMP Algorithm}
			\begin{algorithmic}[1]
            \label{alg:cosamp}
				\Statex Input: $m\times n$ measurement matrix $\Phi$, $m$-dimensional measurement $y$, sparsity level $k$.
                \Statex Output: approximate solution $x_j$
				\State Set $j = 0$
				\State Set the initial solution $x_0 = 0$
				\State Set the current sample $ z = y $
				\State Set the initial support $S = \emptyset$
				\Repeat

				\State Set $j = j + 1$
				\State Form signal proxy $u = \Phi^Tz$
				\State Identify large signal components $\Omega = \text{supp}(u_{(2k)})$
				\State Merge supports $S = \Omega \bigcup \text{supp}(x_{j-1})$
                \State Estimate signal using least-squares method
                $\quad \quad \quad \quad \quad \quad x'_{|S} = \Phi_S^\dagger y, \quad x'_{|S^c} = 0$
                \State Prune the signal to obtain the next approximation
                $\quad \quad  \quad \quad \quad x_j = x'_{(k)}$
                \State Update current sample $z = y - \Phi x_j$
				\Until{halting criterion is true}

			\end{algorithmic}
		\end{algorithm}



\subsection{Combinatorial Algorithms}
These algorithms were mostly developed by the theoretical computer science community, in many cases predating the compressed sensing literature.  There are two major types of algorithms: group testing and data stream sketches. In group testing \cite{gilbert2007}\cite{iwen2010}, there are $n$ items represented by an $n$-dimensional vector $x$, of which there are $k$ anomalous items we would like to identify. The values of elements of $x$ are nonzero if they correspond to anomalous items and zero otherwise. The problem is to design a collection of tests, resulting in the measurement $y = \Phi x$, where the element of matrix $\phi_{i,j}$ indicates $j$th item is included in the $i$th test. The goal is to recover $k$-sparse vector $x$ using the least number of tests, which is essentially a sparse signal recover problem in CS.

An example of data stream sketches \cite{cormode2009, muthukrishnan2005} is to identify the most frequent source or destination IP addresses passing through a network device. Let $x_i$ denote the number of times IP address $i$ is encountered. The vector $x$ is sparse since the network device is likely to see a small portion of IP address space. Directly storing $x_i$ is infeasible since the index set $\{i\}$ is too large ($2^{32}$, since IP address has 32 bits). Instead, we store a sketch defined by $y = \Phi x$, where $\Phi$ is an $m\times n$-dimensional matrix with $m\ll n$. Since $y$ has a linear relationship with $x$, $y$ is incrementally updated each time a packet arrives. The goal is to recover $x$ from the sketch $y$, which again is a sparse signal recovery problem in CS.

    \section{Variations of CS}
		\label{section:extensions}
		In this section, we describe three variations to CS: 1) CS for multiple measurement vector signals, 2) CS for analogy signals, and 3) CS for matrix completion.
\subsection{Multiple Measurement Vectors (MMV)}
Before proceeding further, we mention there is a method used in sensor networks that is related to MMV called distributed compressed sensing or joint sparse signal recovery. We will describe that method in Section VII.B of this paper. In MMV problems, we deal with $l$ correlated sparse signals, which share the same support (the index set of nonzero coefficients). Instead of recovering each signal $x_i$ independently, we would like to recover the signals jointly by exploiting their common sparse support. Let $X = (x_1,x_2,...x_l)$ denote the $n\times l$ matrix representing the signals, and $\Lambda = supp(X)$ denote the sparse support. The CS problem for MMV can be formulated as
        \begin{equation}
		\label{eqn:p_5}
			\text{($P_5$)}  \quad  \underset{X}{\text{min}}  \|X|_1 \text{   subject to } Y = \Phi X
		\end{equation}
where $Y$ is the $m\times l$ measurement matrix, In the above, we assume $X$ is already in a sparse representation to simplify the presentation. It is straightforward to extend the results to the cases where $X$ is not in a sparse representation.

A necessary and sufficient condition to recover signal using CS is given by \cite{davies2012}
        \begin{equation}
		\label{eqn:mmv_cond_1}
			\text{spark}(A) > 2|\text{supp}(X)| - \text{rank}(X) + 1.
		\end{equation}
Assume the signals are $k$ sparse, then $|\text{supp}(X)| = \text{rank}(X) = k$. For an $m\times l$ matrix $A$, spark($A$)$ \leq m + 1$. So, the necessary and sufficient condition for signal recovery becomes
        \begin{equation}
		\label{eqn:mmv_cond_2}
			m > k.
		\end{equation}

Recall that for a single measurement signal, the necessary and sufficient condition for signal recovery is $m \geq 2k$, referring to (\ref{eqn:spark_meas}). Thus MMV CS reduces the number of measurements by half compared with single measurement vector (SMV) CS.

\begin{figure}[htb!]
\label{fig:aic}
		\setlength{\unitlength}{0.13in}
		\centering

		\begin{picture}(35,9)
			\put(4,4.5){\circle{2}}
			\put(8,3){\framebox(7,3)}
			\put(8.5,4.5) {Analog Filter}
			\put(10.5,3.3) {$h(t)$}
			\put(4,8){\vector(0,-1){2.4}}
			\put(0.5,4.5){\vector(1,0){2.4}}
			\put(5,4.5){\vector(1,0){3}}
			\put(15,4.5){\vector(1,0){3}}
			\put(21.5,4.5){\vector(1,0){3}}

			\put(18,4.5){\qbezier(0.5,0.5)(1,3)(3,3.5)}
			\put(18.65,5.5){\vector(-1,-3){0.2}}

			\put(21.5,4.5){\line(-1,1){3}}
			\put(0.3,5.5) {$x(t)$}
			\put(23,5.5) {$y[m]$}
			\put(19,3) {$mM$}
			\put(4.5,8){$p_{c}(t)\epsilon \left \{ -1,1 \right \}$}
			\put(4.7,3.8){\line(-1,1){1.4}}
			\put(4.7,5.2){\line(-1,-1){1.4}}
			\end{picture}
	\hspace{1in}\parbox{6in}{\caption{Pseudo-random demodulation for AIC.\label{fig:aic}}}
\end{figure}
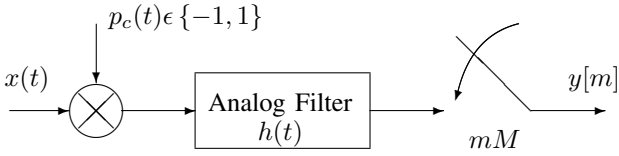
\subsection{Analog-to-Information Conversion (AIC)}
	\label{subsec:aic}
Today, digital signal processing is prevalent. Analog-to-digital converters (ADC) are used to convert analog signals to digital signals. ADC requires sampling at Nyquist rate. The ever-pressing demands from applications, such as ultra-wide-band communications and radar signal detection, are pushing the performance of ADCs toward their physical limits. In \cite{kirolos2006, laska2007, mishali2011}, CS was used to implement an analog-to-information converter. AIC was shown to be particularly effective for wide-band signals that are sparse in frequency domain. We provide an overview of AIC below.

Assume analog sign $x(t)$ is sparse in an orthogonal basis ($\Psi = \{\psi_1(t), \psi_2(t)..., \psi_n(t)\}$) and is expressed by
		\begin{equation}
		\label{eqn:aic_rep}
			x(t) = \sum_{i = 1}^n \theta_i \psi_i(t).
		\end{equation}

AIC is composed of three components, a wide-band pseudo-random signal demolulator $p_c(t)$, a filter $h(t)$ (typically a low-pass filter), and a low-rate analog-to-digital converter (ADC), referring to Figure \ref{fig:aic}. The input of AIC is the signal $x(t)$. The output is the low-rate measurement $y_i$, given by
		\begin{equation}
		\label{eqn:aic_meas1}
			y_i = \int_{-\infty}^{\infty} x(\tau)p_c(\tau)h(t - \tau)\,\mathrm{d} \tau |_{t = m\Delta}
		\end{equation}
where $\Delta$ is the sampling period. Substituting (\ref{eqn:aic_rep}) into (\ref{eqn:aic_meas1}), we have
		\begin{equation}
		\label{eqn:aic_meas2}
			y_i = \sum_{j = 1}^n \theta_j \int_{-\infty}^{\infty} \psi_j(\tau)p_c(\tau)h(i\Delta - \tau)\,\mathrm{d}\tau.
		\end{equation}
The equivalent measurement matrix is given by
		\begin{equation}
		\label{eqn:aic_meas_matrix}
			\phi_{i,j} = \int_{-\infty}^{\infty} \psi_j(\tau)p_c(\tau)h(i\Delta - \tau)\,\mathrm{d}\tau.
		\end{equation}

So the compressed sensing with analog signals can be formulated as
		\begin{equation}
		\label{eqn:p_4}
			\text{($P_3$)}  \quad  \underset{\theta}{\text{min}}  \|\theta\|_1 \quad\text{subject to} \quad y = \Phi\theta
		\end{equation}
which is exactly the same formulation as $P_1$ with $\Psi$ set to identity matrix and the $x$ vector replaced by the $\theta$ vector.

In \cite{laska2007}, simulation was performed using a 200 MHz carrier modulated by a 100 MHz signal. AIC was shown to be able to successfully recover the signal at a sampling rate of one sixth of the Nyquist rate.
\subsection{Matrix Completion}
In matrix completion \cite{candes2009, candes2010, keshavan2010, keshavan2010a}, we seek to recover a low-rank matrix from a small, noisy sample of its elements. Matrix completion has a wide range of applications in collaborative filtering, machine learning, control, remote sensing, computer vision, etc. We first describe the notations used in this subsection. Let $X \in R^{n_1 \times n_2}$ denote the matrix of interest with singular values $\{\sigma_k\}$. Let $\|X\|_{*}$ denote the nuclear norm, which is the $l_1$ norm of the singular value vector. We assume that the samples are randomly selected without replacement, and there is no row and column that is not sampled, since in such case the matrix completion is impossible. In the following, we first describe exact matrix completion \cite{candes2009, keshavan2010}, and then briefly cover noisy matrix completion \cite{candes2010,  keshavan2010a}.

Suppose $M \in R^{n_1 \times n_2}$ is the matrix we want to complete, and we have a small, exact sample of its elements $M_{i,j}, (i,j) \in \Omega$, where $\Omega$ is a subset of indices of $M$. The sampling operator $\mathcal{P}_{\Omega}$ is given by
		\begin{equation}
		\label{eqn:mc_sample}
			[\mathcal{P}_{\Omega}]_{i,j} = \begin{cases} X_{i,j} & \text{if } (i,j) \in \Omega \\
                0       & \text{otherwise}.
   \end{cases}
		\end{equation}

For matrix completion to be possible, we need to impose the following condition on $M$. Let the singular value decomposition (SVD) of $M$ be
		\begin{equation}
		\label{eqn:mc_svd}
			M = \sum_k \sigma_k u_k v_k^*
		\end{equation}
where $\sigma_1\geq\sigma_2\geq...\geq\sigma_r$ are the singular values, and $u_1, u_2,..., u_r \in R^{n_1}$ and $v_1, v_2,..., v_r \in R^{n_2}$ are singular vectors. The condition we impose is as follows
		\begin{equation}
		\label{eqn:mc_cond}
			\|u_k\|_{\infty} \leq \sqrt{\mu_B/n_1},\quad \|v_k\|_{\infty} \leq \sqrt{\mu_B/n_2}
		\end{equation}
for some $\mu_B \geq 1$, where $\|x\|_{\infty} \doteq \text{max}_i |x_i|$. In other words, we impose the condition that the singular vectors are not spiky and are sufficiently spread.

In principal, we can recover the matrix by solving the following minimization problem
		\begin{equation}
		\label{eqn:mc_min_rank}
			\text{min  rank}(X)  \quad  \text{subject to} \quad \mathcal{P}_{\Omega}(X) = \mathcal{P}_{\Omega}(M).
		\end{equation}
However, the above problem is NP-hard \cite{chistov1984}. An alternative is the convex relaxation as follows \cite{candes2009, candes2010}.
		\begin{equation}
		\label{eqn:mc_min_nu_norm}
			\text{min} \quad \|X\|_*  \quad  \text{subject to} \quad \mathcal{P}_{\Omega}(X) = \mathcal{P}_{\Omega}(M).
		\end{equation}
The above nuclear-norm minimization is the tightest convex relaxation of the rank minimization problem \cite{candes2009, candes2010}.

Let $M \in R^{n_1 \times n_2}$ denote a matrix of rank $r = O(1)$ obeying (\ref{eqn:mc_cond}), and $n \doteq \text{max}(n_1, n_2)$. We observe $m$ elements of $M$ with indices uniformly randomly sampled. In \cite{candes2010}, it is shown that for a constant $c$, if the following condition is satisfied
		\begin{equation}
		\label{eqn:mc_min_sample}
			m \geq c\mu_B^4 n \text{log}^2 n
		\end{equation}
then $M$ is the unique solution to (\ref{eqn:mc_min_nu_norm}) with probability of at least $1 - n^{-3}$. In other words, with high probability, nuclear-norm minimization can perform matrix completion without error, using a sample set of size $O(n \text{log}^2 n)$. In addition, if we scale the constant in (\ref{eqn:mc_min_sample}) as $c = c\beta$, the success probability becomes $1 - n^{\beta}$.

An $n_1 \times n_2$ matrix with rank $r$ has $r(n_1 + n_2 -r)$ degrees of freedom (DoF), which is equal to the number of parameters in the SVD. When $r$ is small, the DoF is much smaller than $n_1n_2$, which is the total number of elements in the matrix. The nuclear-norm minimization can recover the matrix using a sample size that exceeds the DoF by a logarithmic factor. To recover the matrix, each row and column must be sampled at least once. It is well know that this occurs when the sample size is of $O(n \text{log} n)$, since it is the same as the coupon collector's problem. Thus, (\ref{eqn:mc_min_sample}) misses the information theoretic limit by a logarithmic factor.

For matrices with all values of the rank, the condition for matrix completion is that the matrix must satisfy the strong incoherence property with a parameter $\mu$, the definition of which is somewhat involved and we refer the reader to the reference \cite{candes2010}. Many matrices obey the strong incoherence property, such as matrices that obey (\ref{eqn:mc_cond}) with $\mu_B = O(1)$, with very few exceptions.

Let $M \in R^{n_1 \times n_2}$ denote a matrix of arbitrary rank $r$ obeying strong incoherence with parameter $\mu$, and $n \doteq \text{max}(n_1, n_2)$. We observe $m$ elements of $M$ with indices uniformly randomly sampled. In \cite{candes2010}, it is shown that for a constant $c$ if the following condition is satisfied
		\begin{equation}
		\label{eqn:mc_min_sample2}
			m \geq c\mu^2 nr \text{log}^6 n
		\end{equation}
then $M$ is the unique solution to (\ref{eqn:mc_min_nu_norm}) with probability of at least $1 - n^{-3}$. In other words, if the matrix is strongly incoherent, with high probability, nuclear-norm minimization can perform matrix completion without error using a sample set of size exceeding the DoF by a logarithmic factor.

In real-world applications, samples always contain noise. Thus, our observation model is given by
		\begin{equation}
		\label{eqn:mc_noise}
			Y_{i,j} = M_{i,j} + Z_{i,j} \text{    or    } \mathcal{P}_{\Omega}(Y) = \mathcal{P}_{\Omega}(M) + \mathcal{P}_{\Omega}(Z)
		\end{equation}
where $Z_{i,j}$ is the noise. We assume that $\|\mathcal{P}_{\Omega}(Y)\|_F \leq \delta$ for some $\delta > 0$. For instance, if $\{Z_{i,j}\}$ is a white noise with standard deviation of $\sigma$, then $\delta^2 \leq (m + 2\sqrt{2m})\sigma^2$ with high probability \cite{candes2010}.

To complete the matrix $M$, we solve the following minimization problem
		\begin{equation}
		\label{eqn:mc_min_nu_norm_noise}
			\text{min} \quad \|X\|_*  \quad  \text{subject to} \quad \|\mathcal{P}_{\Omega}(X - Y)\|_F \leq \delta.
		\end{equation}
In words, we seek the matrix that has the minimum nuclear-norm and is consistent with the data. The above minimization problem can be solved by the FPC algorithm \cite{ma2011} as follows
		\begin{equation}
		\label{eqn:mc_min_nu_norm_noise2}
			\text{min} \quad \frac{1}{2}\|\mathcal{P}_{\Omega}(X - Y)\|_F^2 + \lambda \|X\|_*
		\end{equation}
for some positive constant of $\lambda$.
We refer the reader to \cite{candes2010,  keshavan2010a} for details.

	\section{Applications of CS in the Physical Layer}
		\label{section:CS physical}
		CS can be used in detecting and estimating sparse physical signals, such as MIMO signals, wide-band cognitive radio signals, and ultra-wide-band (UWB) signals, etc. The details are provided below.
\subsection{MIMO Signals}
Channel state information (CSI) is essential for coherent communication over multi-antenna (MIMO) channels. Convention holds that the MIMO channel exhibits rich multi-path behavior and the number of degrees of freedom is proportional to the dimension of the signal space. However, in practice, the impulse responses of MIMO channel actually are dominated by a relatively small number of dominant paths. This is especially true with large bandwidth, long signaling duration, or large number of antennas \cite{czink2007, yan2007}. Because of this sparsity in the multi-path signals, CS can be used to improve the performance in channel estimation.

Consider a MIMO channel with $N_T$ transmitters and $N_R$ receivers. Assume that channel has two-sided bandwidth of $W$ and the signaling has a duration of $T$. Let $s(t)$ denote the $N_T$-dimensional transmitted signal, $\tilde{s}(f)$ its element-wise Fourier transform, $h(t, f)$ the time-varying frequency response matrix, which has a dimension of $N_R\times N_T$. Without the noise, the received signal is given by \cite{bajwa2010}
	\begin{equation}
	\label{eqn:MIMO_rcvd}
		x(t) = \int_{-W/2}^{w/2}h(t,f)\tilde{s}(f)e^{j2\pi ft}\mathrm{d}f.
	\end{equation}

For a multi-path channel, the frequency response is the summation of the contributions from all the paths
	\begin{equation}
	\label{eqn:MIMO_physical}
		h(t,f) = \sum_{n=1}^{N_P}\beta_n a_R(\theta_{R,n})a^H_T(\theta_{T,n})e^{j2\pi \nu_n t}e^{j2\pi \tau_n f}
	\end{equation}
where $N_P$ denotes the number of paths. For path $n$, $\beta_n$ denotes the complex path gain, $\theta_{R,n}$ the angle of arrival (AoA) at the receiver, $\theta_{T,n}$ the angle of departure (AoD) at the transmitter, $\nu_n$ the Doppler shift, and $\tau_n$ the relative delay. The $N_R$-dimensional vector $a_R(\theta_{R,n})$ is array response vector at the receiver. The $N_T$-dimensional vector $a_T(\theta_{T,n})$ is array steering vector at the transmitter, and the superscript $H$ denotes matrix conjugate transpose. We assume that the maximum delay is $\tau_{max}$, then $\tau \in [0, \tau_{max}]$. Also, the two-sided Doppler spread is in the range $\nu \in [-\nu_{max}/2, \nu_{max}/2]$. The maximum antenna angular spread is assumed at the critical antenna spacing ($d = \lambda /2$), thus $(\theta_{R,n}, \theta_{T,n}) \in [-1/2, 1/2]\times [-1/2, 1/2]$ in the normalized unit. It is also assumed that the channel is both time-selective ($T\nu_{max} \leq 1$) and frequency-selective ($W\tau_{max} \leq 1$).

The physical model expressed by (\ref{eqn:MIMO_physical}) is nonlinear and hard to analyze. However, it can be well-approximated by a linear model known as a virtual channel model\cite{sayeed2002,sayeed2003}. The virtual model approximates the physical model by uniformly sampling the physical parameter space $[\beta_n, \theta_{R,n}, \theta_{T,n}, \tau_n, \nu_n]$ at a resolution of $(\Delta\theta_{R,n}, \Delta\theta_{T,n}, \Delta\tau_n, \Delta\nu_n) = (1/N_R, 1/N_T, 1/W, 1/T)$. The approximate channel response is given by
	\begin{eqnarray}
	\label{eqn:MIMO_virtual1}
		h(t,f) \simeq \sum_{i=1}^{N_R}\sum_{k=1}^{N_T}\sum_{l=0}^{L-1}\sum_{m=-M}^{M}h_v(i,k,l,m) \nonumber\\
		a_R(i/N_R)a^H_T(k/N_T)e^{j2\pi \frac{m}{T}t}e^{j2\pi \frac{l}{W}f}
	\end{eqnarray}
	\begin{equation}
	\label{eqn:MIMO_virtual2}
		h_v(i,k,l,m) \simeq \sum_{n \in \text{[sampling point]}}\beta_n
	\end{equation}
where the summation in (\ref{eqn:MIMO_virtual2}) is over all paths that contribute to the sampling point, and a phase and attenuation factor has been absorbed in $\beta_n$. In (\ref{eqn:MIMO_virtual1}), $N_R, N_T, L = \lceil W\tau_{max}\rceil + 1$, and $M = \lceil T\nu_{max} /2\rceil$ denote the maximum numbers of resolvable AoAs, AoDs, delays, and one-sided Doppler shifts. Basically, the virtual model characterizes the physical channel using the matrix $h_v$, which has a dimension of $D = N_R\times N_T\times L\times (2M + 1)$.

For sparse MIMO channels, the number of nonzero elements $d$ in the matrix $h_v$ is far fewer than $D$. We call such channel $d$-sparse. Since the virtual model is linear, we can express the received training signal as
	\begin{equation}
	\label{eqn:MIMO_train}
		y_{tr} = \Phi H_v
	\end{equation}
where $H_v$ is a $D$-dimensional column vector that contains all the elements in $h_v$, ordered according to the index set $(i,k,l,m)$, and $\Phi$ is an $M\times D$-dimensional measurement matrix, which is a function of transmitted training signal and array steering and response vectors. We can formulate a CS problem as follows
	\begin{equation}
	\label{eqn:MIMO_CS}
		H_v = \underset{H_v}{\text{arg min}}\|H_v\|_1 \quad \text{subject to} \quad y_{tr} = \Phi H_v.
	\end{equation}

The requirement for the measurement is that it is a uniformly random sampling of the signal in the domain of $(i,k,l,m)$. It is shown in \cite{bajwa2010} that with high probability of success, roughly $d$ instead of $D$ measurements are enough to recover the $D$-dimensional channel vector $H_v$, which provides significant savings in training resources consumed, especially for sparse-MIMO channels. The details are given in \cite{bajwa2010}. Similar results are also reported in \cite{taubock2008, haupt2010} for RF signals and in \cite{berger2009} for underwater acoustic signals.
\subsection{Wide-Band Cognitive Radio Signals}
Dynamic spectrum access (DSA) is an emerging approach to solve today's radio spectrum scarcity problem. Key to DSA is the cognitive radio (CR) that can sense the environment and adjust its transmitting behavior accordingly to not cause interference to other primary users of the frequency. Thus, spectrum sensing is a critical function of CR. However, wide-band spectrum sensing faces hard challenges. There are two major approaches to do wide-band spectrum sensing. First, we can use a bank of tunable narrow-band filters to search narrow-bands one by one. The challenge with this approach is that a large number of filters need to be used, leading to high hardware cost and complexity. Second, we can use a single RF front-end and use DSP to search the narrow-bands. The challenge in this approach is that very high sampling rate and processing speed are required for wide-band signals. CS can be used to overcome the challenges mentioned above. Today, a small portion of the wireless spectrum is heavily used while the rest is partially or rarely used {fcc2002}. Thus, the spectrum signal is sparse and CS is applicable. In this subsection, we introduce three approaches of applying CS to the spectrum sensing problem.
\subsubsection{Spectrum Sensing: A Digital Approach}
In this approach, the signal is first converted to the digital domain, and then CS is performed on the digital signal. Let $x(t)$ denote the signal sensed by CR, $B$ the frequency range of the wide-band, $F$ the set of frequency bands currently used by other users. Typically, $|F|\ll B$\cite{sahai2005}, indicating that $x(t)$ is sparse in the frequency domain and thus CS is applicable. So, instead of sampling at Nyquist rate $f_N$, we can sample at a much slower rate roughly around $|F|f_N/B$. In \cite{tian2007}, the CS formulation of spectrum sensing was proposed as
	\begin{equation}
	\label{eqn:CR_CS}
		f = \underset{f}{\text{arg min}}  \|f\|_1 \text{   subject to } x_t = Sx(t) = SF^{-1} f
	\end{equation}
where $f$ is signal representation in the frequency domain, $F^{-1}$ is the inverse Fourier transformation, and $S$ is a reduced-rate sampling matrix operating at a rate close to $|F|f_N/B$, and $x_t$ is the reduced rate measurement. Simulation results show that good signal recovery can be achieved at 50\% Nyquist rate\cite{sahai2005}.
\subsubsection{Spectrum Sensing: An Analog Approach}
In this approach, CS is directly performed on the analog signal\cite{yu2008}, which has the advantage of saving the ADC resources, especially in cases where the sampling rate is high. The implementation is similar to that described in Section~\ref{subsec:aic}. A parallel bank of filters are used to acquire measurements $y_i$. To reduce the number of filers required, which is equal to the number of measurements $M$ and can be potentially large, each filter samples time-windowed segments of signal. Let $N_F$ denote the number of filters required, and $N_S$ denote the number of segments each filter acquires. As long as $N_F N_S = M$, the measurement is sufficient. Simulations were performed for an OFDM-based CR system with 256 sub-carriers where only 10 carriers are simultaneously active. The results showed that a CS system with 8-10 filters can perform spectrum sensing at 20/256 of the Nyquist rate \cite{yu2008}.
\subsubsection{Spectrum Sensing: A Cooperative Approach}
The performance of CS-based spectrum sensing can be negatively impacted by the channel fading and the noise. To overcome such problems, a cooperative spectrum sensing scheme based on CS was proposed in \cite{tian2008}. In this scheme, the assumptions are that there are $J$ CRs and $I$ active primary users. The entire frequency range is divided into $F$ non-overlapping narrow-bands $\{B_i\}_{i=1}^F$. In the sensing period, all CRs remain silent and cooperatively perform spectrum sensing. The received signal at $j$th CR is
	\begin{equation}
	\label{eqn:CR_coop_rcvd}
		x_j(t) = \sum_{i = 1}^I h_{i,j}(t)*s_i(t) + n_j(t)
	\end{equation}
where $h_{i,j}$ is the channel impulse response, $s_i(t)$ is the transmitted signal from primary user $i$, $*$ denotes convolution, and $n_j(t)$ denotes noise at the receiver~$j$. We take discrete Fourier transform on $x_j(t)$ and obtain
	\begin{equation}
	\label{eqn:CR_coop_dft}
		\tilde{x}_j(f) = \sum_{i = 1}^I \tilde{h}_{i,j}(f)\tilde{s}_i(f) + \tilde{n}_j(f).
	\end{equation}

In this scheme, spectrum detection is possible even when the channel impulse responses are unknown. If $\tilde{x}_j(f) \neq 0$, then some primary user is using the channel unless the channel suffers from deep fades. Since it is unlikely that all the CR suffer from deep fades at the same time, cooperative spectrum sensing is much more robust than individual sensing.

Cooperative spectrum sensing is carried out in two steps. In the first step, compressed spectrum sensing is carried out at each individual CR using the approach in \cite{tian2007}. In addition, the $j$th CR maintains a binary $F$-dimensional occupation vector $u_j$, where $u_{j,i} = 1$ if the frequency band is sensed to be occupied and $u_i = 0$ otherwise. In the second step, CRs in the one-hop neighborhood exchange the occupation vectors and then update their own occupation vectors using the technique of average consensus \cite{xiao2007} as follows
	\begin{equation}
	\label{eqn:CR_coop_update}
		u_j(t + 1) = u_j(t) + \sum_{k \in N_j}w_{j,k}(u_k(t) - u_j(t))
	\end{equation}
where $N_j$ is the neighborhood of $j$th CR, and $w_{j,k}$ is the weight associated with edge $(j,k)$, the selection rules of which are described in \cite{xiao2007}. With proper selection rules, it can be shown that
	\begin{equation}
	\label{eqn:CR_coop_update}
		\lim_{t \rightarrow \infty} u_j(t) = \frac{1}{J} \sum_{k = 1}^{J} u_k(0).
	\end{equation}

In other words, the frequency occupation vectors of the CRs in the neighborhood all reach the same value that is the average of their initial values. The $j$th CR can decide the frequency band $i$ is occupied if $u_{j,i} \geq 1/J$, or a majority rule can be used, i.e., the frequency band $i$ is considered occupied if $u_{j,i} \geq 1/2$. Simulation results showed that the spectrum sensing performance improves with the number of CRs involved in the cooperative sensing and the average consensus converges fast (in a few iterations) \cite{tian2007}.
\subsection{Ultra-Wide-Band (UWB) Signals}
UWB communications is a promising technology for low-power, high-bandwidth wireless communications. In UWB, an ultra-short pulse, on the order of nanoseconds, is used as the elementary signal to carry information. The advantages of UWB are: 1) The implementation of the transmitter is simple because of the use of base-band signaling. 2) UWB has little impact on other narrow-band signals on the same frequency range, since its power spreads out on the broad frequency range. However, one of the challenges for UWB is that it requires extremely high sampling rate (several GHz) to digitize UWB signals based on the Nyquist rate, leading to high cost in hardware. Since UWB signals are sparse in the time domain \cite{paredes2007}, we can apply CS, which provides an effective solution to this problem by requiring much lower sampling rate. There are a number of recent papers in this area, and we provide two examples below.
\subsubsection{UWB Channel Estimation}
In reference \cite{paredes2007}, CS was applied to UWB multi-path channel estimation. The received multi-path UWB signal can be expressed as
	\begin{equation}
	\label{eqn:uwb_received}
		x(t) = p(t)*h(t) = \sum_{i = 1}^L \theta_i p(t - \tau_i)
	\end{equation}
where $p(t)$ is the ultra-short pulse used to carry information, and $h(t)$ is the impulse response of the UWB channel.  Typically a Gaussian pulse or its derivatives can be used as UWB pulse, i.e., $p(t) = p_n(t)e^{- t^2 / 2 \sigma^2}$, where $p_n(t)$ is a polynomial of degree $n$, and $\sigma$ represents the width of the signal. The impulse response of the channel can be expressed as
	\begin{equation}
	\label{eqn:uwb_channel}
		h(t) = \sum_{i = 1}^L \theta_i \delta(t - \tau_i)
	\end{equation}
where $\delta( )$ is the Dirac delta function, $\theta_i$ and $\tau_i$ are the gain and the delay of the $i$th-path received signal, and $L$ is the total number of propagation paths.

Typically, the number of the channel parameters ($\theta_i$ and $\tau_i$) is on the order of $10^3$. However, most of the paths carry negligible energy and can be ignored. In other words, the channel parameters are sparse and CS can be used to estimate them. Two approaches were proposed in \cite{paredes2007} for UWB channel estimation using CS. One is correlator-based and the other is rake-receiver-based. It was shown that CS-based approaches outperform the traditional detector using only 30\% of ADC resources.
\subsubsection{UWB Echo Signal Detecion}
In reference \cite{shi2008}, the authors proposed to use CS to detect UWB radar echo signals. Let $s(t)$ denote the transmitted signal and $\tau$ as the Nyquist sampling interval of the echo signal. All time-shifted versions of $s(t)$ constitute a redundant dictionary as follows
	\begin{equation}
	\label{eqn:uwb_dict}
		\Psi = \{\psi_i(t) = s(t - i\tau), i \in [1, 2,...,n]\}.
	\end{equation}
The received signal is sparse in such dictionary and can be expressed as
	\begin{equation}
	\label{eqn:uwb_rep}
		x(t) = \sum_{i = 1}^k \theta_i \psi_i(t)
	\end{equation}
where $k$ is the number of target echoes and represents the sparsity of the signal, and $\theta_i$ indicates the amplitude of $i$th target echo. Equation (\ref{eqn:uwb_rep}) has the exact same form as (\ref{eqn:aic_rep}) and standard methods of CS can be used to detect the UWB echo signals. It was shown in \cite{shi2008} that the sampling rate can be reduced to only about 10\% of the Nyquist rate.

    \section{CS Applications as Erasure Code and in the MAC Layer}
		\label{section:CS_in_MAC}
		In the following, we describe CS applications as erasure code and in the MAC layer.
\subsection{CS as Erasure Code}
Many physical phenomena are compressible or sparse in some domains. For example, virtual images are sparse in the wavelet domain and sound signals are sparse in the frequency domain. The conventional approach is to use source coding to compress the signal first, and then use erasure coding for protection against the missing data caused by the noisy wireless channel. Let $x$ denote the $n$-dimensional signal. Let $S$ and $E$ denote the $m\times n$ source coding matrix and $l\times m$ erasure coding matrix, respectively. If the signal is $k$-sparse in some domain, $m$ is close to $k$. If the expected probability of missing data is $p$, then $l = m / (1-p)$. The transmitted signal is $ESx$, and the received signal is $CESx$, where the linear operator $C$ models the channel. $C$ is a sub-matrix of the identity matrix $I_l$ with $e$ rows deleted, $e$ being the number of erasures. If $e \leq l - m$, the decoding at the receiver will be successful; otherwise the data can not be decoded and is discarded.

CS can be used as an effective erasure coding method. In \cite{charbiwala2010}, CS was applied in wireless sensor networks as erasure code. At the source $l$ measurements of the signal are generated by random projections $y = \Phi x$ and sent out, where $\Phi$ is a $l\times n$-dimensional random matrix. The received signal is $y' = Cy$, with $e$ measurements erased. Suppose each measurement carries its serial number, we know where the erasure occurred and therefore the matrix $C$. At the receiver, the standard CS procedure is carried out
	\begin{equation}
	\label{eqn:erasure}
		x = \underset{x}{\text{arg min}}\|\Psi x\|_1 \text{    subject to  } y' = C\Phi x.
	\end{equation}
where $\Psi$ is the representation, under which $x$ is sparse. Since erasure occurs randomly, $C\Phi$ is still a random matrix. CS performs data compression and erasure coding in one stroke. Information about the signal is spread out among $l$ measurements, of which $m$ measurements are expected to be correctly received, with $m$ on the same order of $k$, the sparsity of the data.

Compared with conventional erasure coding methods, CS has similar compression performance but two outstanding advantages: 1) CS allows graceful degradation of the reconstruction error when the amount of missing data exceeds the designed redundancy, whereas the conventional coding methods do not. Specifically, if $e \leq l - m$, conventional decoding can not recover the data at all.  However, if RIP holds, according to (\ref{eqn:RIP-bound}), CS can still recover partial data, with an error no larger than that of the approximate signal, which keeps the largest $l - e$ elements of the sparse signal and sets the rest of the elements to zero. 2) In terms of energy consumed in the processing, performing CS erasure coding is 2.5 times better than performing local source coding and 3 times better than sending raw samples \cite{charbiwala2010}.
\subsection{CS in On-Off Random Access Channels}
In \cite{fletcher2009}, a connection was made between CS and on-off random access channels. In an on-off random multiple access channel, there are $N$ users communicating simultaneously to a single receiver through a channel with $n$ degrees of freedom. Each user transmits with probability $\lambda$. Typically, $\lambda N < n \ll N$. User $i$ is assigned as codeword an $n$-dimensional vector $\phi_i$. The signal at the receiver from user $i$ is $\phi_ix_i$, where $x_i$ is a nonzero complex scalar if the user is active and zero otherwise. The total signal at the receiver is given by
    \begin{equation}
	\label{eqn:on_off_meas}
		y = \sum_{i = 1}^N \phi_ix_i + w = \Phi x +w
	\end{equation}
where $w$ is the noise, $x = [x_1, x_2,...,x_N]$, and $\Phi = [\phi_1, \phi_2,...\phi_N]$ is the codebook. The active user set is defined by
    \begin{equation}
	\label{eqn:on_off_active_usr}
		\Omega = \{i : x_i \neq 0 \}.
	\end{equation}
The goal of the receiver is to estimate $\Omega$. Since $|\Omega| \ll n$, we have a sparse signal detection problem. A formulation in terms of CS is as follows
    \begin{equation}
	\label{eqn:on_off_cs}
		\hat{x} = \underset{x}{\text{arg min  }} \mu \|x\|_1 + \|y - \Phi x\|^2_2
	\end{equation}
where $\mu > 0$ is an algorithm parameter that weights the importance of sparsity in $\hat{x}$.

It was shown in \cite{fletcher2009} that the CS-based algorithms perform better than single-user detection in terms of the number of measurements required to recover the signal, and have some near-far resistance. At high signal-to-noise ratio (SNR), CS-based algorithms perform worse than the optimal maximum likelihood detection. However, CS-based algorithms are computationally efficient, whereas the optimal maximum likelihood detection is not computationally feasible. 

	\section{Applications of CS in the Network Layer}
		\label{section:CS in WSN}
		As mentioned before, most natural phenomena are sparse in some domains, and CS could be effective in wireless sensor networks (WSN) that monitor such phenomena. In \cite{bajwa2006}, CS was used to gather data in a single-hop WSN. Sensors transmit random projections of their data simultaneously in a phase-synchronized channel. The base station receives the summation of the randomly projected data, which constitutes a CS measurement. The $l_1$-norm minimization is performed at the base station to recover the sensory data. An overview of CS' potential applications in WSN was provided in \cite{haupt2008}. In the following, we provide a few detailed examples. We first present a representative CS-based sensory data collection scheme and then present three approaches to reduce the measurement costs: 1) using the joint-sparsity in data to reduce the number of measurements, 2) using sparse random projection to reduce number of measurements, and 3) using data routing to reduce the cost of data transport. We compare the performance of various data collection schemes at the end of the section.
\subsection{Compressive Data Gathering (CDG) in WSN}
In \cite{luo2009}, CS was used for data collection in WSN. In the following, we describe the proposed data collection scheme, and how it handles the abnormal sensor readings.
\subsubsection{The CS-Based Data Collection Scheme}
Consider a large-scale sensor network with $n$ nodes, each of which holds a value $x_i$. Assume that a shortest-path spanning tree rooted at the base station is built, and that the base station and sensors agree to the seeds for random number generation. To collect the data, $m$ measurements are taken. For measurement $j$, the data transmissions start from the leaf-nodes of the spanning tree and work in rounds. In each round, the children send their data to the parent in the tree. Specifically, in the $j$th round, each leaf-node sends the random projection of its data ($y_{j,i} = \phi_{j,i}x_i$) to its parent in the tree. The parent node $l$ collects data sent by all its children and compute an partial measurement as follows
	\begin{equation}
	\label{eqn:cdg_meas}
		y_{j,l} = \sum_{i \in N_l}y_{j,i} + \phi_{j,l}x_l
	\end{equation}
where $N_l$ denotes the children of node $l$. The parent node in turn sends its partial measurement to its own parent. This continues until the base station receives the partial measurements from its children and compute the complete measurement $y_j = \sum_{i = 1}^n \phi_{j,i}x_i$.
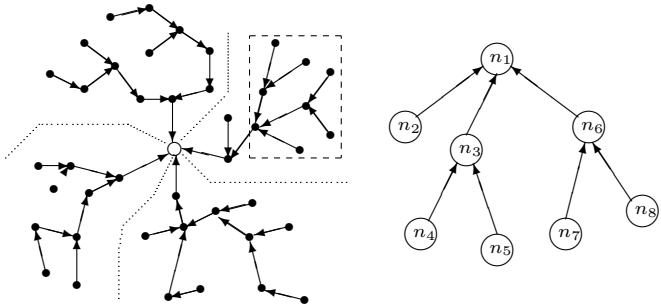
\begin{figure}[htb!]

		\centering
		
		\centering
		\setlength{\unitlength}{.08in}
		\begin{picture}(12,27)

			\put(-10,23){\circle*{0.5}}
		         \put(-10,23){\vector(4,- 3){2}}
		         \put(-10,20){\circle*{0.5}}
		         \put(-10,20){\vector(4,3){2}}
		         \put(-8,21.5){\circle*{0.5}}
		         \put(-8,21.5){\vector(3,-4){1.6}}
		         \put(-8.3,24.7){\circle*{0.5}}
		         \put(-8.3,24.7){\vector(4,1){2.5}}
		         \put(-5.75,25.35){\circle*{0.5}}
		         \put(-5.75,25.35){\vector(4,- 3){2}}
		         \put(-5.75,22.35){\circle*{0.5}}
		         \put(-5.75,22.35){\vector(4,3){2}}
		         \put(-3.75,23.85){\circle*{0.5}}
		         \put(-3.75,23.85){\vector(3,-2){1.8}}
		         \put(-12.24,21){\circle*{0.5}}
			\put(-12.24,21){\vector(2,-1){2}}
			\put(-4.24,19.34){\circle*{0.5}}
		         \put(-6.34,19.34){\circle*{0.5}}
			\put(-6.34,19.34){\vector(1,0){2}}
			\put(-1.8,22.5){\circle*{0.5}}
			\put(-1.8,22.5){\vector(0,-1){2.5}}
			\put(-1.8,20){\circle*{0.5}}
			\put(-1.8,20){\vector(-4,-1){2.5}}
			\put(-4.24,19.34){\vector(0,-1){2.8}}

			\put(2.5,23){\circle*{0.5}}
			\put(2.5,23){\vector(-1,-4){0.75}}
			\put(1.7,19.8){\circle*{0.5}}
			\put(1.7,19.8){\vector(-1,-4){0.5}}
			\put(1.2,17.55){\circle*{0.5}}
			\put(1.2,17.6){\vector(-3,-4){1.7}}
			\put(-0.6,18.1){\circle*{0.5}}
			\put(-0.6,18.1){\vector(0,-1){2.6}}
			\put(-0.6,15.4){\circle*{0.5}}
			\put(-0.6,15.4){\vector(-4,1){3}}%
			\put(4.5,21.8){\circle*{0.5}}
			\put(4.5,21.8){\vector(-3, -2){2.7}}
			\put(6.1,21.1){\circle*{0.5}}
			\put(6.1,21){\vector(-3, -4){1.5}}
			\put(6.1,17){\circle*{0.5}}
			\put(6.1,17){\vector(-3, 4){1.5}}
			\put(4.5,18.9){\circle*{0.5}}
			\put(4.5,19){\vector(-2, -1){3}}
			\put(4.1,16){\circle*{0.5}}
			\put(4.1,16){\vector(-2, 1){2.8}}

			\put(-13,15){\circle*{0.5}}
			\put(-13,15){\vector(1,0){2}}
			\put(-10.9,15){\circle*{0.5}}
			\put(-10.9,15){\vector(4,-1){3}}
			\put(-7.7,14.2){\circle*{0.5}}
			\put(-7.7,14.2){\vector(2,1){3.2}}%
			\put(-12,13.5){\circle*{0.5}}
			\put(-12,13.5){\vector(2,3){1}}
			\put(-13.2,11){\circle*{0.5}}
			\put(-13.2,11){\vector(4,-1){2.5}}
			\put(-12.5,8){\circle*{0.5}}
			\put(-12.5,8){\vector(-1,4){0.7}}
			\put(-9.7,13.2){\circle*{0.5}}
			\put(-9.7,13.2){\vector(2,1){1.9}}
			\put(-10.5,10.3){\circle*{0.5}}
			\put(-10.5,10.3){\vector(1,4){0.7}}
			\put(-10.5,7.2){\circle*{0.5}}
			\put(-10.5,7.2){\vector(0,1){3}}

			\put(-4,13){\circle*{0.5}}
			\put(-4,13){\vector(0,1){2.7}}
			\put(-3.5,11){\circle*{0.5}}
			\put(-3.5,11){\vector(-1,4){0.45}}
			\put(-6,10.4){\circle*{0.5}}
			\put(-6,10.4){\vector(4,1){2.3}}
			\put(-4.5,6.4){\circle*{0.5}}
			\put(-4.5,6.4){\vector(1,4){1.1}}
			\put(-2.4,6.9){\circle*{0.5}}
			\put(-2.4,6.9){\vector(-4,-1){2}}
			\put(-1.4,12){\circle*{0.5}}
			\put(-1.4,12){\vector(-2,-1){2}}
			\put(1,12.55){\circle*{0.5}}
			\put(1,12.55){\vector(-4,-1){2.2}}
			\put(0.8,10.3){\circle*{0.5}}
			\put(0.8,10.3){\vector(-3,2){2}}
			\put(3.4,11){\circle*{0.5}}
			\put(3.4,11){\vector(-4,-1){2.3}}
			\put(1.6,7){\circle*{0.5}}
			\put(1.6,7){\vector(-1,4){0.75}}
			\put(4.4,6.2){\circle*{0.5}}
			\put(4.4,6.2){\vector(-4,1){2.6}}

			\put(0.8,15.5){\dashbox{0.4}(6,8)}
			\put(-4.1,16.1){\circle{0.8}}
			\multiput(-3.7, 16.7)(1,1){3} {\line(1,0){0.1}}
			\multiput(-3.5, 16.9)(1,1){3} {\line(1,0){0.1}}
			\multiput(-3.3, 17.1)(1,1){3} {\line(1,0){0.1}}
			\multiput(-3.1, 17.3)(1,1){3} {\line(1,0){0.1}}
			\multiput(-2.9, 17.5)(1,1){3} {\line(1,0){0.1}}
			
			\multiput(-0.6, 19.8)(0,1){4} {\line(0,1){0.1}}
			\multiput(-0.6, 20)(0,1){4} {\line(0,1){0.1}}
			\multiput(-0.6, 20.2)(0,1){4} {\line(0,1){0.1}}
			\multiput(-0.6, 20.4)(0,1){4} {\line(0,1){0.1}}
			\multiput(-0.6, 20.6)(0,1){4} {\line(0,1){0.1}}
			
			\multiput(-4.8, 16.5)(-2,1){1} {\line(1,0){0.1}}
			\multiput(-5, 16.6)(-2,1){1} {\line(1,0){0.1}}
			\multiput(-5.2, 16.7)(-2,1){1} {\line(1,0){0.1}}
			\multiput(-5.4, 16.8)(-2,1){1} {\line(1,0){0.1}}
			\multiput(-5.6, 16.9)(-2,1){1} {\line(1,0){0.1}}
			\multiput(-5.8, 17)(-2,1){1} {\line(1,0){0.1}}
			\multiput(-6, 17.1)(-2,1){1} {\line(1,0){0.1}}
			\multiput(-6.2, 17.2)(-2,1){1} {\line(1,0){0.1}}
			\multiput(-6.4, 17.3)(-2,1){1} {\line(1,0){0.1}}
			\multiput(-6.6, 17.4)(-2,1){1} {\line(1,0){0.1}}
			\multiput(-6.8, 17.5)(-2,1){1} {\line(1,0){0.1}}
			\multiput(-7, 17.6)(-2,1){1} {\line(1,0){0.1}}
			\multiput(-7.2, 17.7)(-2,1){1} {\line(1,0){0.1}}
			
			\multiput(-7.2, 17.7)(-1,0){6} {\line(1,0){0.1}}
			\multiput(-7.5, 17.7)(-1,0){6} {\line(1,0){0.1}}
			\multiput(-7.8, 17.7)(-1,0){6} {\line(1,0){0.1}}
			
			\multiput(-13, 17.7)(-2,-3){1} {\line(1,0){0.1}}
			\multiput(-13.2, 17.5)(-2,-3){1} {\line(1,0){0.1}}
			\multiput(-13.4, 17.3)(-2,-3){1} {\line(1,0){0.1}}
			\multiput(-13.6, 17.1)(-2,-3){1} {\line(1,0){0.1}}
			\multiput(-13.8, 16.9)(-2,-3){1} {\line(1,0){0.1}}
			\multiput(-14, 16.7)(-2,-3){1} {\line(1,0){0.1}}
			\multiput(-14.2, 16.5)(-2,-3){1} {\line(1,0){0.1}}
			\multiput(-14.4, 16.3)(-2,-3){1} {\line(1,0){0.1}}
			\multiput(-14.6, 16.1)(-2,-3){1} {\line(1,0){0.1}}
			\multiput(-14.8, 15.9)(-2,-3){1} {\line(1,0){0.1}}
			\multiput(-15, 15.7)(-2,-3){1} {\line(1,0){0.1}}
			\multiput(-15.2, 15.5)(-2,-3){1} {\line(1,0){0.1}}
			
			\multiput(-4.2, 15.7)(-2,-3){1} {\line(1,0){0.1}}
			\multiput(-4.3, 15.5)(-2,-3){1} {\line(1,0){0.1}}
			\multiput(-4.4, 15.3)(-2,-3){1} {\line(1,0){0.1}}
			\multiput(-4.5, 15.1)(-2,-3){1} {\line(1,0){0.1}}
			\multiput(-4.6, 14.9)(-2,-3){1} {\line(1,0){0.1}}
			\multiput(-4.7, 14.7)(-2,-3){1} {\line(1,0){0.1}}
			\multiput(-4.8, 14.5)(-2,-3){1} {\line(1,0){0.1}}
			\multiput(-4.9, 14.3)(-2,-3){1} {\line(1,0){0.1}}
			\multiput(-5, 14.1)(-2,-3){1} {\line(1,0){0.1}}
			\multiput(-5.1, 13.9)(-2,-3){1} {\line(1,0){0.1}}
			\multiput(-5.2, 13.7)(-2,-3){1} {\line(1,0){0.1}}
			\multiput(-5.3, 13.5)(-2,-3){1} {\line(1,0){0.1}}
			\multiput(-5.4, 13.3)(-2,-3){1} {\line(1,0){0.1}}
			\multiput(-5.5, 13.1)(-2,-3){1} {\line(1,0){0.1}}
			\multiput(-5.7, 12.9)(-2,-3){1} {\line(1,0){0.1}}
			\multiput(-5.9, 12.7)(-2,-3){1} {\line(1,0){0.1}}
			\multiput(-6.1, 12.5)(-2,-3){1} {\line(1,0){0.1}}
			\multiput(-6.3, 12.3)(-2,-3){1} {\line(1,0){0.1}}
			\multiput(-6.5, 12.1)(-2,-3){1} {\line(1,0){0.1}}
			\multiput(-6.7, 11.9)(-2,-3){1} {\line(1,0){0.1}}
			\multiput(-6.9, 11.7)(-2,-3){1} {\line(1,0){0.1}}
			\multiput(-7.1, 11.5)(-2,-3){1} {\line(1,0){0.1}}
			\multiput(-7.3, 11.3)(-2,-3){1} {\line(1,0){0.1}}
			\multiput(-7.5, 11.1)(-2,-3){1} {\line(1,0){0.1}}
			\multiput(-7.6, 10.8)(-2,-3){1} {\line(1,0){0.1}}
			\multiput(-7.6, 10.5)(-2,-3){1} {\line(1,0){0.1}}
			\multiput(-7.7, 10.2)(-2,-3){1} {\line(1,0){0.1}}
			\multiput(-7.7, 9.9)(-2,-3){1} {\line(1,0){0.1}}
			\multiput(-7.8, 9.6)(-2,-3){1} {\line(1,0){0.1}}
			\multiput(-7.8, 9.3)(-2,-3){1} {\line(1,0){0.1}}
			\multiput(-7.8, 9)(-2,-3){1} {\line(1,0){0.1}}
			\multiput(-7.8, 8.7)(-2,-3){1} {\line(1,0){0.1}}
			\multiput(-7.8, 8.4)(-2,-3){1} {\line(1,0){0.1}}
			\multiput(-7.8, 8.1)(-2,-3){1} {\line(1,0){0.1}}
			\multiput(-7.8, 7.8)(-2,-3){1} {\line(1,0){0.1}}
			\multiput(-7.8, 7.5)(-2,-3){1} {\line(1,0){0.1}}
			\multiput(-7.8, 7.2)(-2,-3){1} {\line(1,0){0.1}}
			\multiput(-7.8, 6.9)(-2,-3){1} {\line(1,0){0.1}}
			\multiput(-7.8, 6.6)(-2,-3){1} {\line(1,0){0.1}}
			\multiput(-7.8, 6.3)(-2,-3){1} {\line(1,0){0.1}}
			
			\multiput(-3.6, 15.7)(2,-3){1} {\line(1,0){0.1}}
			\multiput(-3.4, 15.5)(2,-3){1} {\line(1,0){0.1}}
			\multiput(-3.2, 15.3)(2,-3){1} {\line(1,0){0.1}}
			\multiput(-3, 15.1)(2,-3){1} {\line(1,0){0.1}}
			\multiput(-2.8, 14.9)(2,-3){1} {\line(1,0){0.1}}
			\multiput(-2.6, 14.7)(2,-3){1} {\line(1,0){0.1}}
			\multiput(-2.4, 14.5)(2,-3){1} {\line(1,0){0.1}}
			\multiput(-2.2, 14.3)(2,-3){1} {\line(1,0){0.1}}
			\multiput(-2, 14.1)(2,-3){1} {\line(1,0){0.1}}
			\multiput(-1.8, 13.9)(2,-3){1} {\line(1,0){0.1}}
			\multiput(-1.5, 13.9)(1,0){9} {\line(1,0){0.1}}
			\multiput(-1.2, 13.9)(1,0){9} {\line(1,0){0.1}}
			\multiput(-0.9, 13.9)(1,0){9} {\line(1,0){0.1}}

			\scriptsize\put(16.5,21.8){$n_1$}
				\put(17,22){\circle{2}}
			
			\put(10.5,17.3){$n_2$}
				\put(11,17.5){\circle{2}}
				\put(11.7,18.2){\vector(4,3){4.4}}
			\put(14.5,15.8){$n_3$}
				\put(15,16){\circle{2}}
				\put(15,17){\vector(1,2){2}}

			\put(22.5,17.3){$n_6$}
				\put(23,17.5){\circle{2}}
				\put(22.3,18.2){\vector(-4,3){4.4}}
				
			\put(16.5,9.3){$n_5$}
				\put(17,9.5){\circle{2}}
				\put(17,10.5){\vector(-1,3){1.55}}
			
			\put(11.5,10.3){$n_4$}
				\put(12,10.5){\circle{2}}
				\put(12.5,11.4){\vector(1,2){1.9}}	
			
			\put(26,11.8){$n_{8} $}
				\put(26.5,12){\circle{2}}
				\put(25.8,12.7){\vector(-2,3){2.55}}

			\put(21,10.3){$n_{7}$}
				\put(21.5,10.5){\circle{2}}
				\put(21.5,11.5){\vector(1,4){1.25}}

		\end{picture}
 	\hspace{2in}\parbox{3in}{\caption{Data collection tree in CDG.\label{fig:cdg}}}
	\end{figure}
An example is given in Figure \ref{fig:cdg}, where the data gathering tree is shown on the left and a portion of the tree is magnified on the right. At the $j$th round,  Node 4 and 5 send their data $\phi_{j,4}x_4$ and $\phi_{j,5}x_5$ to node 3. Node 3 computes the partial measurement $\phi_{j,3}x_3 + \phi_{j,4}x_4 + \phi_{j,5}x_5$, and sends it to node 1. Similar procedure is performed for other nodes.

The base station is able to recover the data by solving the CS problem below
	\begin{equation}
	\label{eqn:cdg_recover}
		x = \underset{x}{\text{arg min}}\|\Psi x\|_1 \quad \text{subject to} \quad \|y - \Phi x\|_2 < \epsilon
	\end{equation}
where $\Psi$ is the basis under which the data is sparse, and $\epsilon$ is the error tolerance.

Using convention methods, the energy consumption in data transmission is very unevenly distributed. The children of the base station are responsible to relaying $O(n)$ pieces of data, causing them to die quickly due to battery depletion. On the other hand, in CDG each node transmit exactly $m$ times, and the energy consumption is perfectly balanced.

One shortcoming of CDG is that the nodes close to the leaves of the network are required to send more pieces of data than that of the conventional method. A leaf-node is required to send one piece of data (its own sensory reading) in the conventional method, but with CS it is required to send $m$ pieces of data. To solve this problem, a hybrid CS scheme was proposed, where if a node sends less than $m$ pieces of data in the conventional method, the conventional method is used, otherwise CS is used\cite{luo2010}.

\subsubsection{Data Recovery with Abnormal Readings}
CS can also be used to recover data with abnormal readings. For example, if a signal is smooth in the time domain, then its representation in the Fourier domain is sparse. If spikes are injected into the time domain signal, then the signal's representation in the frequency domain is no longer sparse. However, we can decompose the signal into two parts
	\begin{equation}
	\label{eqn:cdg_abnormal}
		x = x_0 + x_1
	\end{equation}
where $x_0$ and $x_1$ represent normal and abnormal parts, respectively. Since the spikes are sparse in the time domain and the normal signal is sparse in the frequency domain, the $l_1$-norm minimization is altered to
	\begin{equation}
	\label{eqn:cdg_recover}
		x = \underset{x}{\text{arg min}}\|\Psi x_0 + I x_1 \|_1 \quad \text{subject to} \|y - \Phi x\|_2 < \epsilon
	\end{equation}
where $\Psi$ is the Fourier transformation matrix and $I$ is the identity matrix. A representation like $\Psi x_0 + I x_1$ is called an over-complete representation. Donoho et al. showed it is feasible to achieve stable recovery of the signal in over-complete representations \cite{donoho2006a}.
\subsection{Distributed Compressed Sensing (DCS) in WSN}
Consider a WSN monitoring a natural phenomenon. Among the signals obtained by the sensors, there are likely both intra-signal and inter-signal correlations. Leveraging these correlations, DCS uses the joint sparsity of the signals to reduce the number of measurements for signal recovery \cite{baron2005, duarte2006}, in a fashion similar to MMV described in Section IV.A, but with some differences in problem formulation. DCS requires no collaboration among the sensors, and provides universal encoder for any jointly sparse signal ensemble. In the following, we first introduce three joint sparsity models and then describe signal recovery algorithms.
\subsubsection{Joint Sparsity Models (JSM)}
Assume the signals obtained by the sensors are $x_j, j = 1,2,...J$, and they are sparse in basis $\Psi$.

\emph{JSM1}: In this model, each signal is composed of a common sparse part and an individual sparse part as follows
\begin{equation}
\label{eqn:dcs_jsm1}
x_j = z_0 + z_j \quad \text{with} \quad  z_0 = \Psi\theta_0,  z_j = \Psi\theta_j
\end{equation}
where the coefficients $\theta_0$ and $\theta_j$ are $k_0$ and $k_j$-sparse, respectively. An example of this model is the temperature signals, which can be decomposed into a global average value plus a value reflecting local variations.

\emph{JSM2}: In this model, all signals share the same index set of nonzero coefficients in the sparse representation, and different signals have different individual coefficients as follows
\begin{equation}
\label{eqn:dcs_jsm2}
x_j = \Psi\theta_j
\end{equation}
where the coefficients $\theta_j$ are $k$-sparse. An example of this model is the image observed by multiple sensors, which has the same sparse wavelet-representation but each sensor senses a different value due to different levels of phase-shift and attenuation.

\emph{JSM3}: This model is an extension of JSM1 in that the common part is no longer sparse in any basis
\begin{equation}
\label{eqn:dcs_jsm3}
x_j = z_0 + z_j \quad \text{with} \quad  z_0 = \Psi\theta_0,  z_j = \Psi\theta_j
\end{equation}
where the coefficients $\theta_j$ are $k_j$-sparse. An example of this model is the signals detected by sensors in the presence of strong noise, which is not sparse in any representation.
\subsubsection{Joint Reconstruction}
To reconstruct the signal, sensor $j$ acquires its $n$-dimensional signal $x_j$, takes random projections of the signal $y_j = \Phi_jx_j$, where $\Phi_j$ is the $m_j\times n$-dimensional measurement matrix, and sends $y_j$ to the base station. After receiving the measurements from all the sensors, the base station starts the reconstruction, which is different for each of the sparsity models and is described separately below.

\emph{JSM1}: To recover the signal, the following linear program is sovled
\begin{eqnarray}
\label{eqn:dcs_jsm1_rcst}
[\theta_0, \theta_1, \theta_2,...\theta_J] = \underset{\theta_0, \theta_1, \theta_2,...\theta_J}{\text{arg min}} \sum_{j = 0}^J \|\theta_j\|_1 \nonumber \\
\text{subject to} \quad y_j = \Phi\Psi\theta_j \quad \forall{j = 0, 1, 2,...,J}.
\end{eqnarray}

\emph{JSM2}: In this model, conventional greedy pursuit algorithms (such as OMP) are modified. Specificly, step 7 in Algorithm 1 is modified so that the index set inserted to $S_j$ includes only the indices of the common support, i.e., the nonzero items having the same indices among all sensors.

\emph{JSM3}: In this model, each sensor's signal is the addition of a common part $z_0$ that is not sparse and a sparse signal $z_j$ called innovation. The alternating common and innovation estimation (ACIE) scheme was proposed to recover signals \cite{duarte2006}. ACIE alternates between two steps: 1) Estimate $z_0$ by treating $z_j$ as noise that can be averaged out. 2) Estimate $z_j$ by subtracting $z_0$ from the signal and using conventional CS recovery techniques.

Extensive simulations have been carried out that demonstrate CS leveraging joint sparsity models can significantly reduce the number of measurements required. For detail, the reader is referred to \cite{baron2005}.
\subsection{Sparse Random Projection (SRP) in WSN}
The measurement matrix $\Phi$ used in conventual CS is dense in the sense that there are few zero elements in each row of $\Phi$. This means each measurement requires data from $O(n)$ sensors, which is expensive to acquire. Fortunately, it was shown in \cite{wang2007} that sparse random projections (SRP) can be used to reduce the cost of measurement, where the measurement matrix is given by
	\begin{equation}
	\label{eqn:wsn_srp}
		\phi_{j,i}  = \left\{
		  \begin{array}{l l}
  			   +1 & \quad \text{with prob. $p = 1/2s$}\\
   			  -1 & \quad \text{with prob. $p = 1/2s$}\\
   			  0 & \quad \text{with prob. $p = 1 - 1/s$}
   		\end{array} \right.
	\end{equation}
where $s$ is a parameter that determines the sparseness of the measurement. In other words, SRP requires data from only $O(n/s)$ instead of $O(n)$ sensors. In order to recover a $k$-sparse signal with high probability, i.e. with probability $1 - n^{-\gamma}$ for some positive $\gamma$, the number of measurements required is given by
	\begin{equation}
	\label{eqn:wsn_meas}
		m = O(sM^2k^2\log n)
	\end{equation}
where $M$ is the peak-to-total energy ratio and is given by
	\begin{equation}
	\label{eqn:wsn_ptr}
		M \equiv \frac{\|x\|_\infty}{\|x\|_2}.
	\end{equation}
$M$ bounds the largest element of the signal. SRP works well only if the signal is not too concentrated in a few elements.

\begin{figure}[htb!]
		\centering
		\setlength{\unitlength}{0.14in}
		\centering

		\begin{picture}(20,15)

			\put(15,9){\circle{1}}
			\put(5,3){\circle{1}}
			\put(7,13){\circle{1}}
			\put(4,8){\circle{1}}
			\put(12,8){\circle{1}}
			\put(9,10){\circle*{1}}
			\put(20,8){\circle{1}}
			\put(15,4){\circle*{1}}
			\put(4,6){\circle*{1}}
			\put(13,13){\circle*{1}}
			\put(9,2){\circle*{1}}
			\put(14,8){\circle*{1}}
			 \put(15,4){\vector(-3,4){2.7}}
 			\put(4,6){\vector(4,1){7.5}}
			 \put(9,2){\vector(1, 2){2.8}}
 			\put(14,8){\vector(-3,0){1.5}}
			 \put(13,13){\vector(-1, -4){1.1}}
			 \put(9,10){\vector(3, -2){2.6}}

		\end{picture}
 	    \hspace{2in}\parbox{3in}{\caption{SRP: the node in the center receives data from a randomly selected subset of nodes (dark ones) in the network.\label{fig:srp}}}
	\end{figure}
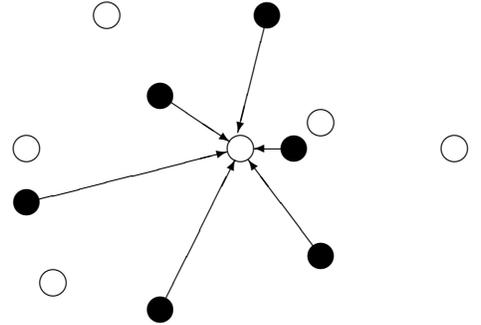

A distributed algorithm for data collection in WSN was proposed in \cite{wang2007}, which works in two steps. In the first step, assuming the base station and sensors agree to the seeds for random number generation, each sensor $i$ generates a random projection $\phi_{j,i}x_i$. If $\phi_{j,i}$ is zero, nothing needs to be done, otherwise the sensor sends the random projection to another sensor chosen randomly, referring to Figure \ref{fig:srp}. Since $\Phi$ is $s$-sparse, each sensor sends $n/s$ pieces of data. In the second step, sensor $j$ waits until it receives $n/s$ pieces of data and then computes a measurement $y_j = \sum_{i = 1}^{n/s}\phi_{j,i}x_i$. The base station sends query to $m$ sensors, and gets $m$ measurements back. As long as $m$ satisfies (\ref{eqn:wsn_meas}), the signal is recovered successfully with high probability.

There is a tradeoff involved in SRP. The larger the value of $s$, the less the cost of data spreading among the sensors, but the higher the cost of the query from the base station. Suppose the average number of hops between two nodes is $h$. The data spreading in SRP requires $hn^2/s$ transmissions. The data query requires $O(hsM^2k^2\log n)$ transmissions. The optimum value of $s$ is given by
	\begin{equation}
	\label{eqn:wsn_s}
		s = O(\frac{n}{Mk\sqrt{\log n}})
	\end{equation}

\begin{table*}[!ht]
\caption {Transport costs for various data collection schemes}
\begin{center}
\begin{tabular}{ l  p{5cm} }
\hline
Data collection schemes & data transport cost ($n$: number of nodes, $k$: data sparsity, $s$: measurement sparsity)   \\ \hline
Conventional data collection                       &  $O(n^{3/2})$  \\
CS Gossip \cite{rabbat2006}                        &  $O(n^2)$         \\
CS Spanning tree   \cite{lee2009a, lee2009b}       &  $O(nk\log{n/k})$     \\
CS Spanning tree + SRP  \cite{lee2009a, lee2009b,wang2007}    &  $O(nk\log{n/k}/s)$ \\ \hline
\end{tabular}
\end{center}
\label{table:data_transport}
\end{table*}
\subsection{Optimizing Data Routing for CS in WSN}
Conventional CS assumes each measurement costs the same. In WSN, the assumption is no longer true. Each measurement is a linear combination of data from a number of different sensors, which entails data transport cost. The design of the data collection scheme has implications on the data transport cost. SRP reduces the cost of measurement by reducing the number of sensors required for each measurement. Another approach to reduce the measurement cost is to optimize data routing. Using this approach, a spatially localized compressed sensing and routing scheme was proposed \cite{lee2009a, lee2009b}, where measurements were formed among adjacent sensors. A shortest-distance spanning tree is used to collect data to the base station.

A cautionary note was raised in \cite{quer2009}, where the authors studied the interplay between routing and signal representation for compressed sensing in WSN. In WSN, each row of the measurement matrix $\Phi$ actually represents a path or route, where the nonzero elements in the row represent the nodes encountered in the path. The condition for good reconstruction quality is that $\Phi$ and $\Psi$ are incoherent, where $\Psi$ is the basis on which the signal is sparse. The authors of \cite{quer2009} showed that the condition is not always met in practice. They used both synthetic and real data sets, and considered a number of popular sparsifying transformations such as DCT, Haar Wavelet, etc. They showed that reconstruction quality of CS for synthetic data is good, but not as good for real data sets, such as Wi-Fi signal strengths, temperature levels, etc. This implies that investigating the interaction between the routing and the sparsifying transformation is still an open problem.
\subsection{Transport Cost of Data Collection Schemes}
To collect all the values using conventional methods, $O(n^{3/2})$ transmissions are required, since there are $n$ nodes and each node has an average distance of $O(n^{1/2})$ hops to the base station. In\cite{rabbat2006}, a randomized gossiping scheme was proposed to collect sensory data using CS in WSN. The technique of average consensus similar to (\ref{eqn:CR_coop_update}) is used. The basic gossip scheme incurs a data transport cost of $O(n^2)$ transmissions for a network of $n$ nodes, since it requires broadcasting data from each sensor to all other sensors.

Next, we consider CS data collection schemes based on the spanning tree. Suppose the signal $x$ is $k$-sparse in some domain. According to (\ref{eqn:condition_rip}), $m$ measurements are sufficient to recover the signal, where $m = ck\log(n/k)$. Using CS to collect data, $O(nm)= O(nk\log{n/k})$ transmissions are required, since $n$ transmissions are required to collect one measurement in a spanning tree, and there are $m$ measurements. Furthermore, if SRP is used, the data transportation cost is reduced by a factor of $s$, where $s$ is the sparsity of the measurement as mentioned in the previous section. When $k \ll n$, using CS can save transmission cost significantly. The transport costs of various data collection schemes are summarized in Table \ref{table:data_transport}.

    \section{Applications of CS in the Application Layer}
		\label{section:CS in Monitoring}
		In this section, we describe CS applications in the application layer, specifically how CS is used in network monitoring. Effective performance monitoring is essential for the operation of large-scale networks. The challenge that conventional monitoring techniques face is that they do not scale well, since only a small portion of a large-scale network can be monitored. CS can be used to overcome this challenge. In the following we describe three approaches of using CS for network performance monitoring: 1) directly applying CS for sparse signal monitoring, and 2) applying CS in the transform domain for network performance monitoring, 3) leveraging spatial-temporal correlations in data to deal with missing data in the CS application.
\subsection{Applying CS Directly}
To monitor the performance of the network, monitors are placed at the nodes to measure the end-to-end performance, such as delay, packet loss, etc. In a network of $n$ links, the path-oriented performance measurements can be expressed by
    \begin{equation}
	\label{eqn:path_measure}
		y = \Phi x
	\end{equation}
where $y$ is the $m$-dimensional path performance metrics vector, $x$ is the $n$-dimensional link performance metrics vector, and $\Phi$ is the $m\times n$-dimensional binary matrix called routing matrix. If path $i$ includes link $j$, then $\phi_{i,j} = 1$, otherwise, $\phi_{i,j} = 0$. The above model applies to some metrics such as delay, but not to others such as packet loss (actually, it can be made to be applicable after a log transformation).

Network administrators are generally interested in identifying a few severely congested links with large delays, compared to which delays of other links are negligible. In this sense, $x$ is sparse. A CS-based scheme was proposed to recover $x$ in \cite{firooz2010}. The problem remaining to be addressed is whether the routing matrix $\Phi$ is a good measurement matrix. The authors of the paper leveraged the fact that routing matrices of bipartite expander   graphs are good measurement matrices for CS \cite{gilbert2010}.

\begin{figure}[htb!]
		\setlength{\unitlength}{0.09in}
		\centering

		\begin{picture}(20,25)
		
			\put(4,4){\circle{2}}
			\put(4,8){\circle{2}}
			\put(4,11.5){\circle{2}}
			\put(3.8,15){\circle{2}}
			\put(4,18.5){\circle{2}}
			\put(4,22){\circle{2}}
			
			\footnotesize\put(3.5,3.8) {$e_6$}
		         \put(5,4){\line(3,5){5.27}}
			\put(5,4){\line(5,2){5.2}}
			
		         \put(3.5,7.8) {$e_5$}
		         \put(5,8){\line(3,5){5.27}}
			\put(5,8){\line(5,2){5.2}}
			
		         \put(3.5,11.3) {$e_4$}
		         \put(5,11.5){\line(3,5){5.2}}
			\put(5,11.5){\line(4,-1){5.2}}
			
		         \put(3.5,14.8) {$e_3$}
		         \put(10.2,20.2){\line(-6,-5){5.6}}
			\put(4.7,15.5){\line(3,-1){5.3}}
			
		         \put(3.5,18.3) {$e_2$}
		         \put(5,18.4){\line(2,-3){5.3}}
			\put(5,18.5){\line(3,-1){5.2}}
			
			\put(3.5,21.8) {$e_1$}
			\put(5,22){\line(1,-3){5.2}}
			\put(5,22){\line(3,-1){5.2}}
			
			\put(11.2,6){\circle{2}}
			\put(11.2,10){\circle{2}}
			\put(11,13.5){\circle{2}}
			\put(11.2,17){\circle{2}}
			\put(11.2,20.5){\circle{2}}
			
			\put(10.7,5.8) {$P_5$}
			\put(10.7,9.8) {$P_4$}
		         \put(10.5,13.3) {$P_3$}
		         \put(10.7,16.8) {$P_2$}
			\put(10.7,20.3) {$P_1$}
			
		\end{picture}
	\hspace{2in}\parbox{3in}{\caption{Bipartite graph with edges on the left and paths on the right.\label{fig:bipartite}}}
\end{figure}
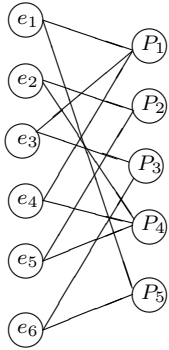

A bipartite graph $G(L,R,E)$ consists of two sets of nodes: the left set $L$ and the right set $R$, referring to Figure \ref{fig:bipartite}. The edges of the graph are only between nodes in $L$ and $R$. The routing matrix can be transformed into a bipartite graph by making $L$ be composed of the links in the network and $R$ be composed of paths in the network. There is a link between node $i$ in $L$ and node $j$ in $R$, only if link $i$ is in path $j$.
A $(s,d,\epsilon)$-expander graph is a bipartite graph $G(L,R,E)$ with left degree $d$ (all nodes in $L$ having $d$ edges), and for any subset $S \in L$ with $|S| \leq s$, the following holds
    \begin{equation}
	\label{eqn:expander}
		|N(S)| \geq (1 - \epsilon)d|S|
	\end{equation}
where $N(S)$ is the set of neighbors of $S$. Parameters $s$ and $\epsilon$ are called expansion factor and error parameter. In other words, in expander graphs $G(L,R,E)$, $L$ is expansive in the sense that any subset $S\in L$ has a neighborhood size proportional to the size of $S$.

It was shown that the bipartite matrix of a $(2s, d, \epsilon)$-expander graph can be used as a good measurement matrix of a $s$-sparse signal \cite{indyk2008, gilbert2010}. Reference \cite{firooz2010} considered the special case where $x$ is 1-sparse. The following CS problem was formulated
    \begin{equation}
	\label{eqn:direct_monitor_cs}
		x = \underset{x}{\text{arg min}}\|x\|_1 \quad \text{subject to} \ y=\Phi x.
	\end{equation}
Let $x^*$ denote the true delay vector. It was shown that if the network graph is a $(2, d, \epsilon)$-expander with $\epsilon \leq 1/4$, then the  following holds \cite{firooz2010}
    \begin{equation}
	\label{eqn:direct_monitor_error}
		\|x - x^*\|_1 \leq c(\epsilon)\|x_c\|_1
	\end{equation}
where $c(\epsilon)$ is a constant dependent on $\epsilon$, and $x_c$ is $x^*$ with the $k$-largest elements removed ($k = 1$ in this case). So, if the true delay vector is 1-sparse, then $x_c = 0$, i.e., the estimation error is zero.

One cautionary note: the above scheme applies only to cases where the network bipartite graphs are expanders. But for some networks, the partite graphs are not expanders, and such scheme is not applicable.
\subsection{Applying CS in the Transform Domain}
A CS-based network monitoring scheme was proposed in \cite{coates2007}. The CS-based scheme provides a scalable monitoring technique that requires measurements on only a few end-to-end paths. Diffusion wavelet \cite{coifman2006} was used as the sparsity-inducing basis. Diffusion wavelet is the generalization of wavelets that provides multi-scale decomposition of functions defined on graphs. For details, the reader is referred to \cite{coifman2006}.

The task is to monitor the performance metrics, such as end-to-end delay or packet loss rate, on a collection of $n_p$ end-to-end paths in a network. The number of paths actually measured $n_s$ is much less than $n_p$. Let $y$ denote the vector of performance values for paths 1 to $n_p$, and $y_s$ denote the vector of performance values of the subset of measured paths. The problem is to estimate $y$, given $y_s$. We can relate $y$ and $y_s$ using the following equation
    \begin{equation}
	\label{eqn:moni_select}
		y_s = Ay
	\end{equation}
where $A$ is the identity matrix with $n_p - n_s$ rows deleted, retaining $n_s$ rows corresponding to measurements $y_s$.

A measurement graph $G(V,E)$ can be constructed as follows. The vertices $V$ of $G$ corresponds to the paths of the physical network. There is an edge between two vertices $v_i$ and $v_j$, if paths $i$ and $j$ share a link in the physical network. A link is assigned a weight $w_{i,j}$ to indicate the degree of correlation of performance metrics between two vertices (two physical paths). We can define the weight such that it is proportional to the fraction of shared physical links in the two paths. Let $L_i$ denote the set of physical links in path $i$. Thus, weight $w_{i,j}$ is given by
    \begin{equation}
	\label{eqn:moni_weigth}
		w_{i,j} = \frac{|L_i \cap L_j|}{|L_i \cup L_j|}.
	\end{equation}

We assign each vertex $v_i$ with the value of performance metric of path $i$. Thus, we obtain a performance metric function $y(V)$ defined on the vertices $V$ of $G$. This function can be represented in diffusion wavelet basis as follows
    \begin{equation}
	\label{eqn:moni_diffusion_basis}
		y = \sum_{i = 1}^n b_i\beta_i = B\beta
	\end{equation}
where $b_i$'s are orthonormal diffusion wavelets defined on the $V$, $\beta_i = y^T b_i$ is the $i$th wavelet coefficient, $B$ is an $n\times n$ matrix composed of $b_i$'s, and $\beta$ is an $n$-dimensional vector composed of $\beta_i$'s.

By proper selection of the diffusion wavelet basis (the reader is referred to \cite{coates2007} for details), we can make $y$'s representation to be sparse, where CS is effective. What is given is $y_s$, which is a subset of path metrics. The goal is to reconstruct $y$, which is the entire set of path metrics. We can formulate a CS problem as follows
    \begin{equation}
	\label{eqn:moni_cs}
		\hat{\beta} = \underset{\beta}{\text{arg min} \|\beta\|_1}\quad\text{subject to}\quad y_s = Ay = AB\beta.
	\end{equation}
The reconstructed path metrics are given by
    \begin{equation}
	\label{eqn:moni_reconst}
		\hat{y} = B\hat{\beta}.
	\end{equation}

The above method was applied to two case studies \cite{coifman2006}. The first case study is monitoring the end-to-end delay in the Abilene network consisted of 11 nodes and 30 unidirectional links. It was shown that it takes only 3 measurements per time step to estimate the mean network end-to-end delay with an error of less than 10\%. The error decreases with more time steps and eventually approaches zero. The second case study is monitoring the bit-error rate (BER) in an all-optical network. The NSF network was used in the simulation, which consists of 14 nodes and 42 unidirectional links. It was shown that even if only 5 monitors are used, BER of more that 60\% lightpaths can be estimated. When 15 monitors are used, BER of more than 90\% lightpaths can be estimated. These results indicate that CS-based network monitoring brings significant savings in measurement resources and costs.
\subsection{Spatial-Temporal Compressed Sensing of Traffic Matrices}
In \cite{zhang2009}, CS was applied to measure traffic matrices. A traffic matrix (TM) specifies the traffic volumes exchanged between the origin and destination pairs in a network during a particular time period. Traffic matrices are important in traffic engineering, capacity planning, and network anomaly detection. The proposed scheme leverages the fact that TMs exhibit pronounced spatial and temporal correlations to deal with the problem of missing data in the measurement. The problem of missing data is frequently encountered in large-scale traffic measurements. CS applies here since TMs are sparse, and CS is effective in dealing with missing data.

However, existing CS algorithms do not perform well on real TMs, and are not flexible to incorporate the desired range of applications. The proposed scheme is called sparsity regularized matrix factorization (SRMF). SRMF finds sparse approximations to TMs, which is augmented by spatial-temporal modeling and local interpolation to obtain high accuracy.
\subsubsection{backgroud}
In this subsection, we provide background knowledge about traffic matrices and singular value decomposition.

{\em Traffic Matrices}: Let $T(i,j;t)$ denote the TM that indicates the volume of traffic in bytes between a source $i$ and a destination $j$ during the time interval $[t, t + \Delta t]$. The TM $T$ is a 3-dimensional array, i.e., $T \in R^n\times R^n\times R^m$, where $n$ is the number of nodes in the network and $m$ is the number of time intervals. Let $x_t$ be the vector constructed by stacking the columns of $T$. Let $X \in R^n\times R^m$ denote the matrix consists of $x(t)$'s from $m$ time intervals. Let $Y$ denote the link traffic load, which is related to $X$ as shown below
    \begin{equation}
	\label{eqn:tm_link}
		Y = AX
	\end{equation}
where $A$ is the routing matrix, whose element $A_{i,j}$ indicates that the $j$-th source-destination flow traverses link $i$. The TM inference problem is to seek the best estimate $\hat{X}$ based on the link measurement $Y$. In fact, reference \cite{zhang2009} includes the generalization to other measurements such as flow-records at the routers, which is omitted here for the easy of the exposition.

{\em Singular Value Decomposition (SVD)}:
With SVD, an $n\times m$ real matrix $X$ is decomposed as
    \begin{equation}
	\label{eqn:svd_x}
		X = U\Sigma V^T
	\end{equation}
where $U$ is an $n\times n$ unitary matrix, i.e., $U^TU = UU^T = I$, $V$ is a $m\times m$ unitary matrix, $V^T$ is the transpose of $V$, and $\Sigma$ is an $n\times m$ diagonal matrix consisting of the singular values $\sigma_i$ of $X$. Singular values are arranged so that $\sigma_i \geq \sigma_{i+1}$. The rank of a matrix is the number of independent rows or columns, which is the same as the number of non-zero singular values.

We can write SVD in the following form
    \begin{equation}
	\label{eqn:svd_x}
		X = U\Sigma V^T = \sum_{i=1}^{\text{min}(n,m)}\sigma_i u_i v_i^T  = \sum_{i=1}^{\text{min}(n,m)}A_i
	\end{equation}
where $u_i$ and $v_i$ are the $i$-th columns of $U$ and $V$, respectively, and matrices $A_i$ are rank-1 by construction. The best rank-$r$ approximation of $X$ with respect to Frobenius norm, i.e., the solution to the following problem
    \begin{equation}
	\label{eqn:min_frob}
		\text{min} \|X - \hat{X}\|_F, \quad \text{subject to  rank}(\hat{X}) \leq r,
	\end{equation}
is given by
    \begin{equation}
	\label{eqn:svd_r}
		X = \sum_{i=1}^{r}A_i
	\end{equation}
where the Frobenius norm is defined by
    \begin{equation}
	\label{eqn:norm_f}
		\|X\| = \sqrt{\sum_{i,j}X_{i,j}^2}.
	\end{equation}
We rewrite the SVD in the following form
    \begin{equation}
	\label{eqn:svd_x2}
		X = U\Sigma V^T = LR^T
	\end{equation}
where $L = U\Sigma^{1/2}$ and $R = V\Sigma^{1/2}$, and we will use this form in the rest of the subsection.

{\em CS of TM}: It was shown in \cite{lakhina2004} that TMs inhibit in a relatively low dimensional space. Thus, we can recover TMs by solving the following problem
    \begin{equation}
	\label{eqn:min_rank1}
		\text{min}\quad\text{
rank}(LR^T) \quad \text{subject to}\quad A(LR^T) = Y.
	\end{equation}
Rank minimization is a non-convex optimization problem and is hard to solve. However, when A holds RIP and the rank of $A$ is less than that of $LR^T$, then (\ref{eqn:min_rank1}) is equivalent to
    \begin{equation}
	\label{eqn:min_rank2}
		\text{min}\quad\|L\|_F^2 + \|R\|_F^2 \quad \text{subject to}\quad A(LR^T) = Y.
	\end{equation}

Since real TM is often only approximately low-rank and the measurements often contain errors, seeking a strictly low-rank solution would likely fail. Thus, we seek a low-rank approximation without strictly enforcing the measurement equations as the following
    \begin{equation}
	\label{eqn:min_rank3}
		\text{min}\quad\|ALR^T - B\|_F^2 + \lambda(\|L\|_F^2 + \|R\|_F^2)
	\end{equation}
where $\lambda$ is a tunable parameter that controls the tradeoff between achieving low rank and the fit to the measurements.

We obtain $L$ and $R$ from \ref{eqn:min_rank3} using an alternative least-squares procedure. First, we initialize $L$ and $R$ randomly. Then, we solve the optimization problem \ref{eqn:min_rank3} by taking one of $L$ and $R$ fixed and the other the optimization variable. We alternate $L$ and $R$'s roles and continue until the convergence, at which time we obtain a solution. The above approach is called sparsity regularized SVD (SRSVD) interpolation.
\subsubsection{Spatial-Temporal Compressed Sensing}
The scheme proposed in \cite{zhang2009} seeks to capture both global and local structures in TMs. TMs are shown to have strong cyclic behavior \cite{eriksson2010}, due to the diurnal or weekly traffic cycles. TMs also exhibit strong spatial structure \cite{roughan2005}. The proposed scheme consists of two components: 1) sparsity regularized matrix factorization (SRMF) to capture global spatial-temporal structures; and 2) a mechanism to incorporating local interpolation.

{\em SRMF:} We seek to leverage the fact that there exist spatial-temporal correlations in TMs, which means that the rows or columns of TMs close to each other are close in values. So, instead of solving (\ref{eqn:min_rank3}), we solve the following
    \begin{eqnarray}
	\label{eqn:min_rank4}
		\text{min}\quad\|ALR^T - B\|_F^2 + \lambda(\|L\|_F^2 + \|R\|_F^2) \nonumber\\
          + \|S(LR^T)\|_F^2 + \|(LR^T)T^T\|_F^2
	\end{eqnarray}
where $S$ and $T$ are spatial and temporal constraint matrices, which express our knowledge about the spatial-temporal structure of the TM. We solve the above optimization problem using the alternative least squares method described earlier. The resulting algorithm constitutes SRMF. Compared to SRSVD, SRMF allows us to express other objectives by selecting different choices of $S$ and $T$.

{\em Selection of $S$ and $T$}: Both SRSVD and SRMF require the specifications of the input ranks of $L$ and $R$. However, it has been shown that SRMF is not sensitive to the input ranks.

We first discuss the choice of $T$. A simple choice of $T$ is $T_p(0, 1, -1)$, which denotes the Toeplitz matrix with its elements defined by
	\begin{equation}
	\label{eqn:tp_mtx}
		T_p{(i,j)}  = \left\{
		  \begin{array}{l l}
  			   +1 & \quad i = j\\
   			  -1 & \quad i = j - 1\\
   			  0 & \quad \text{otherwise}
   		\end{array} \right.
	\end{equation}
This temporal constraint matrix reflects the fact that TMs at adjacent times are often similar in values. The matrix $XT^T$ is the matrix of differences between immediately adjacent elements of $X$. Minimizing $\|(LR^T)T^T\|_F^2$ implies seeking a solution having similar temporally adjacent values.

Next, we discuss the choice of $S$. We choose $S$ by first deriving an initial estimate $\hat{X}$ using a simple interpolation algorithm, and then choose $S$ according to the similarity between rows of $\hat{X}$, as described in the following:
\begin{itemize}
\item  Deriving $\hat{X}$: We derive $\hat{X}$ as follows
    \begin{equation}
	\label{eqn:drv_x}
		\hat{X} = X_{base}*(1 - M) + D * M
	\end{equation}
where $*$ represents element-wise product, $D$ contains the direct measurement, and $M$ indicates which TM elements are being measured directly and is given by
	\begin{equation}
	\label{eqn:missing}
		M_{i,j}  = \left\{
		  \begin{array}{l l}
  			   0 & \quad \text{if $X_{i,j}$ is missing}\\
   			   1 & \quad \text{otherwise}
   		\end{array} \right.
	\end{equation}
In other words, we use direct measurements where available, and interpolate using $X_{base}$ otherwise.
\item  Selecting $S$: There are a number of possible ways to select $S$ based on $\hat{X}$. A general method is the following: 1) We construct a weighted graph $G$, where each node corresponds to a row of $\hat{X}$, and each edge weight indicates a certain similarity measure between two nodes. 2) We set $S$ to be the normalized Laplacian matrix \cite{chung1996} of graph $G$, which behaves as a differencing operator on $G$ and induces sparsity by eliminating redundancy between similar nodes.
    In particular, for each row $i$, we find the $K$ most similar rows $j_k \neq i$. We preform linear regression to obtain a set of weights $w_k$ such that the row $i$ is best approximated by a linear combination of rows $j_k$ as in the following
    \begin{equation}
	\label{eqn:wghts_lr}
		\hat{X}_{i, *} \simeq \sum_{k=1}^K w_k \hat{X}_{j_k, *}.
	\end{equation}
    We select $S$ using the following formula
    \begin{equation}
	\label{eqn:s_select}
		S_{i, i} = 1 \quad\text{and}\quad S_{i,j_k} = - w_k.
	\end{equation}
\end{itemize}

Finally we need to scale $S$ and $T$ so that $\|S(LR^T)\|_F^2, \|(LR^T)T^T\|_F^2$ and $\|ALR^T - B\|_F^2$ have similar order of magnitude. Otherwise, some terms would overwhelm others in the optimization. The scaling is performed as follows
    \begin{equation}
	\label{eqn:scl_mtx}
		\|S\hat{X}\|_F = 0.1\sqrt{\lambda}\|B\|_F \quad\text{and}\quad \|\hat{X}T^T\|_F = \sqrt{\lambda}\|B\|_F
	\end{equation}
where $\sqrt{\lambda}\|B\|_F$ represents the level of approximation error $\|ALR^T - B\|_F^2$ we would tolerate. We make $\|S\hat{X}\|_F$ smaller than $\|\hat{X}T^T\|_F$, because the temporal model acquired through domain knowledge is more reliable than the spatial one. It was shown in \cite{zhang2009} that such scaling provides good performace across a wide-range of scenarios and the performance is not sensitive to the selection of $\lambda$.

\subsubsection{Combining Global and Local Methods}
To take advantage of the local structure of TMs, we use SRMF as a prior and augment it with a local interpolation procedure. Two hybrid algorithms are proposed: SRMF+KNN($k$-nearest neighbor) and SRSVD+KNN.
With SRMF+KNN, first, we compute the SRMF interpolation of $X$, with the result being $X_{SRMF}$. For each missing data point $X_{i,j}$. If any of the elements $X_{i,j-3},...,X_{i,j+3}$ are present, then we use them to approximate $X_{i,j}$. Otherwise, we use $X_{SRMF i,j}$. With SRSVD, the only difference is that we use $X_{SRSVD}$ rather than $X_{SRMF}$. The performance of SRMF+KNN is superior to that of SRSVD+KNN.

The above scheme was evaluated using real TM data across a wide-range of missing-values scenarios, from random to structured, and from low-level to high-level missing-values scenarios. It performs significantly better than commonly-used alternative techniques, such as low-rank approximation schemes and KNN local interpolation schemes.

    \section{Conclusion}
		\label{section:Conclusion}
		In this paper, we introduce CS theory and algorithms and its applications in communications networks. We show that CS has varied applications ranging from the physical layer through the application layer, and in many cases brings performance gains on the order of 10X. We believe this is just the beginning of CS applications in communications networks, and the future will see even more fruitful applications of CS in our field.

\end{document}